\RequirePackage[final]{graphicx}
\documentclass[aps, superscriptaddress, prx, twocolumn]{revtex4-1}
\usepackage[utf8]{inputenc}
\usepackage{amsmath}
\usepackage{amssymb}
\usepackage{amsfonts}
\usepackage{mathtools}
\usepackage{fixme}
\usepackage{color}
\usepackage{braket}
\usepackage{dsfont}
\usepackage{hyperref}

\usepackage{tikz}
\usetikzlibrary{calc}

\usetikzlibrary{positioning,arrows}
\usetikzlibrary{decorations}
\usetikzlibrary{decorations.pathmorphing,positioning}
\usetikzlibrary{decorations.markings}

\tikzset
{
hole/.style 		= { draw = black, postaction = { decorate }, decoration = { markings, mark = at position .55 with { \arrow[black]{ triangle 45} } } },
spinwave_11/.style 		= { draw = cyan, postaction = { decorate }, decoration = { markings, mark = at position .5 with { \arrow[cyan]{ triangle 45} } } },
particle_small_arrow/.style 		= { draw = black, postaction = { decorate }, decoration = { markings, mark = at position .5 with { \arrow[scale = 0.6, black]{ triangle 45} } } },
spinwave_12/.style 		= { draw = cyan, postaction = { decorate }, decoration = { markings, mark = at position .25 with { \arrow[cyan, >=triangle 45]{<} }, mark = at position .75 with { \arrow[cyan, >=triangle 45]{>} } } },
spinwave_21/.style 		= { draw = cyan, postaction = { decorate }, decoration = { markings, mark = at position .25 with { \arrow[cyan, >=triangle 45]{>} }, mark = at position .75 with { \arrow[cyan, >=triangle 45]{<} } } },
spinwave_total/.style 			= { decorate, decoration = {snake, amplitude = 2pt, segment length = 7pt } }
}

\tikzset{
  ncbar angle/.initial=90,
  ncbar/.style={
    to path=(\tikztostart)
    -- ($(\tikztostart)!#1!\pgfkeysvalueof{/tikz/ncbar angle}:(\tikztotarget)$)
    -- ($(\tikztotarget)!($(\tikztostart)!#1!\pgfkeysvalueof{/tikz/ncbar angle}:(\tikztotarget)$)!\pgfkeysvalueof{/tikz/ncbar angle}:(\tikztostart)$)
    -- (\tikztotarget)
  },
  ncbar/.default=0.5cm,
}

\tikzset{square left brace/.style={ncbar=0.1cm}}
\tikzset{square right brace/.style={ncbar=-0.1cm}}

\definecolor{myred}{RGB}{214,26,70}
\definecolor{myreddark}{RGB}{76,8,38}
\definecolor{myblue}{RGB}{35,106,185}
\definecolor{mybluedark}{RGB}{19,56,99}
\definecolor{mybluebright}{RGB}{225,236,249}

\def\te{{\rm e}}
\def\ba{{\bf a}}

\def\bd{{\bf d}}
\def\bk{{\bf k}}
\def\bp{{\bf p}}
\def\bq{{\bf q}}
\def\br{{\bf r}}

\def\bQ{{\bf Q}}
\def\bK{{\bf K}}

\def\bdelta{{\pmb \delta}}

\def\pa{\partial}
\def\nn{\nonumber}

\def\AF{{ \rm AF }}

\begin{document}
\title{Spatial structure of magnetic polarons in strongly interacting antiferromagnets}
\date{\today}

\author{K. \ K. \ Nielsen}
\affiliation{Center for Complex Quantum Systems, Department of Physics and Astronomy, Aarhus University, Ny Munkegade 120, DK-8000 Aarhus C, Denmark}
\author{M. \ A. \ Bastarrachea-Magnani}
\affiliation{Center for Complex Quantum Systems, Department of Physics and Astronomy, Aarhus University, Ny Munkegade 120, DK-8000 Aarhus C, Denmark}
\affiliation{Departamento de F{\'i}sica, Universidad Aut{\'o}noma Metropolitana-Iztapalapa, Av. San Rafael Atlixco 186, CP 09340 CDMX, Mexico}
\author{T. \ Pohl}
\affiliation{Center for Complex Quantum Systems, Department of Physics and Astronomy, Aarhus University, Ny Munkegade 120, DK-8000 Aarhus C, Denmark}
\author{G. \ M. \ Bruun}
\affiliation{Center for Complex Quantum Systems, Department of Physics and Astronomy, Aarhus University, Ny Munkegade 120, DK-8000 Aarhus C, Denmark}
\affiliation{Shenzhen Institute for Quantum Science and Engineering and Department of Physics, Southern University of Science and Technology, Shenzhen 518055, China}

\begin{abstract} 
The properties of mobile impurities in quantum magnets are fundamental for our understanding of strongly correlated materials and may play a key role in the physics of high-temperature superconductivity. Hereby, the motion of hole-like defects through an antiferromagnet has been of particular importance. It creates magnetic frustrations that lead to the formation of a quasiparticle, whose complex structure continues to pose substantial challenges to theory and numerical simulations. In this article, we develop a non-perturbative theoretical approach to describe the microscopic properties of such magnetic polarons. Based on the self-consistent Born approximation, which is provenly accurate in the strong-coupling regime, we obtain a complete description of the polaron wave function by solving a set of Dyson-like equations that permit to compute relevant spin-hole correlation functions. We apply this new method to analyze the spatial structure of magnetic polarons in the strongly interacting regime and find qualitative differences from predictions of previously applied truncation schemes. Our calculations reveal a remarkably high spatial symmetry of the polaronic magnetization cloud and a surprising misalignment between its orientation and the polaron crystal momentum. The developed framework opens up an approach to the microscopic properties of doped quantum magnets and will enable detailed analyses of ongoing experiments based on cold-atom quantum simulations of the Fermi-Hubbard model. 
\end{abstract}

\maketitle

\section{Introduction} \label{sec.introduction}
The Fermi-Hubbard Hamiltonian is a paradigmatic model in condensed matter physics, introduced to describe the behavior of electrons in a solid~\cite{Hubbard1963}. It supports a remarkably broad spectrum of quantum phases of matter, and it is believed to capture the essential phenomenology of strongly correlated materials including the cuprates \cite{Lee2006}. Yet, the Fermi-Hubbard model has proven extremely difficult to analyze and continues to challenge theoretical and numerical efforts for more than four decades \cite{SchmittRink1988,Shraiman1988,Kane1989,Martinez1991,Liu1991,Emery1987, Schrieffer1988, Dagotto1994,Anderson1987,Bonca1989,Hasegawa1989,Dagotto1990,Sachdev1989,Trugman1990,Boninsegni1992,Brunner2000,Mishchenko2001,Blomquist2020,White2001,Zhu2014,Wang2021,Chen2021,Bulaevskii1968,Brinkman1970,Trugman1988,Manousakis2007,Grusdt2018,Grusdt2018_2,Grusdt2019,Bohrdt2019,Bohrdt2021,Soriano2020}. An important case emerges close to half filling where each lattice site is occupied by one fermion. Then, strong on-site particle repulsion leads to the build-up of antiferromagnetic order of the spins of the fermions, which competes with the delocalization of holes that can be present in the lattice \cite{Chao1977,Hirsch1985,Izyumov1997}. This results in a buildup of magnetic frustrations around such holes and the formation of quasiparticles, termed magnetic polarons \cite{SchmittRink1988,Shraiman1988,Kane1989,Martinez1991,Liu1991}. The emerging magnetic dressing cloud induces effective interactions between two such holes that have been conjectured to provide a mechanism for high-temperature superconductivity \cite{Emery1987,Schrieffer1988,Dagotto1994}. Understanding and characterizing magnetic polarons has therefore been of key interest for many decades. 

Owing to the shear complexity of the problem, only a few theoretical approaches have been applied under different conditions and with varying success. This includes exact diagonalization for small system sizes \cite{Bonca1989,Hasegawa1989,Dagotto1990,Wang2021}, mean field approaches \cite{Anderson1987}, and variational calculations \cite{Sachdev1989,Trugman1990}. One has also analysed the string-excitations caused by defect motion through the magnet \cite{Bulaevskii1968,Brinkman1970,Trugman1988,Manousakis2007,Grusdt2018,Grusdt2018_2,Grusdt2019,Bohrdt2019}, and employed numerical techniques such as Monte-Carlo simulations \cite{Boninsegni1992,Brunner2000,Mishchenko2001,Blomquist2020}, machine learning methods \cite{Bohrdt2019} as well as renormalization group techniques \cite{White2001,Zhu2014,Wang2021,Chen2021}. 

\begin{figure}[t!]
\begin{center}
\includegraphics[width=1\columnwidth]{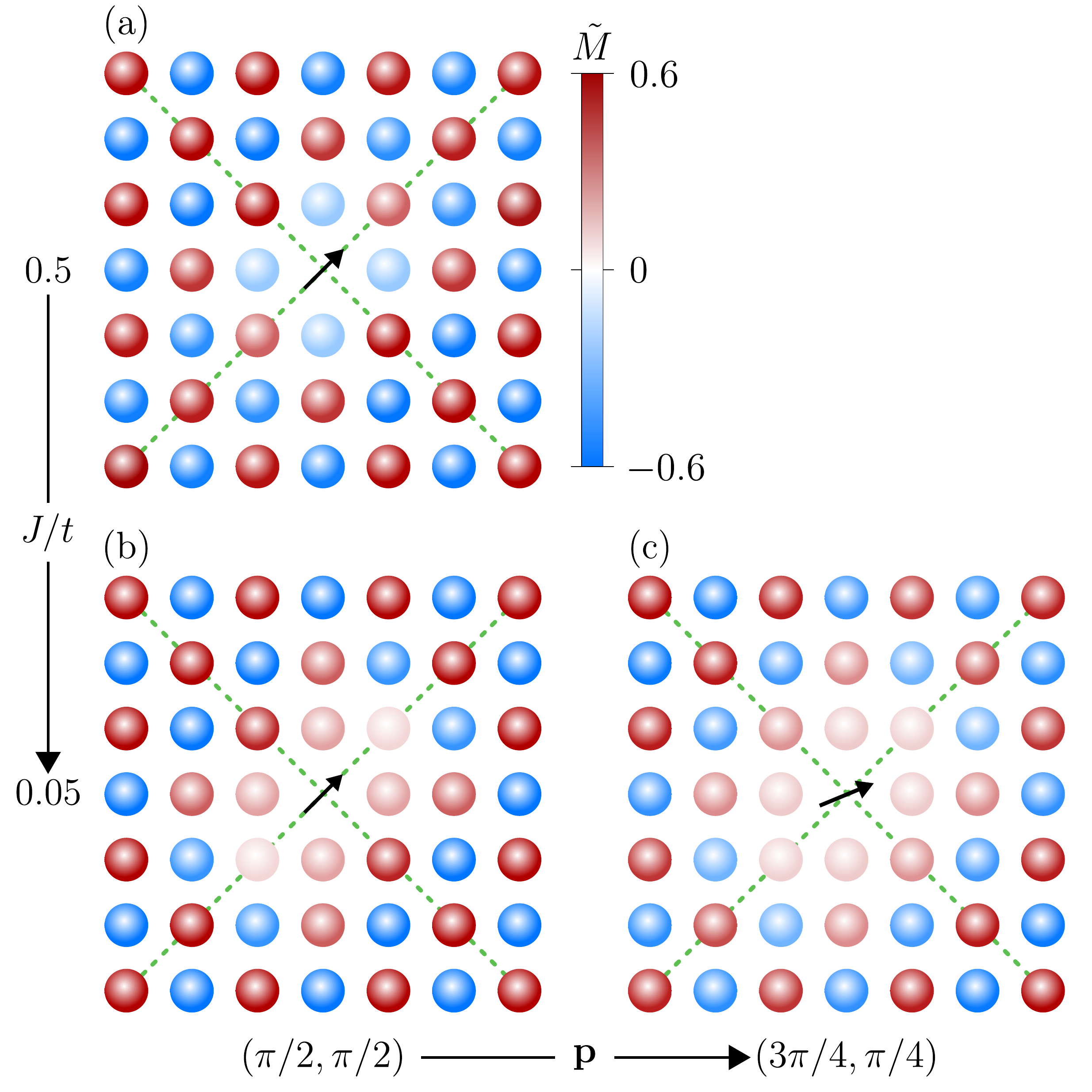}
\end{center}
\caption{Magnetization around a hole in a Heisenberg antiferromagnet in a 16 by 16 square lattice for different interaction strengths and crystal momenta along the magnetic Brillouin zone. As the hopping amplitude $t$ becomes large compared to the spin-spin coupling $J$, moving from panels (a) to (b), the magnetic frustrations around the hole increase in size and magnitude and the magnetic order even flips sign at the nearest neighbor sites. The dashed lines indicate the reflection symmetries of the magnetization, which along with a $C_2$ rotation symmetry means that the dressing cloud has the remarkably high symmetry group $C_{2v}$. As a consequence, the orientation of the dressing cloud is in general misaligned with the crystal momentum leading to a non-trivial behavior.}
\label{fig.magnetization_fig1} 
\end{figure} 

Recent experimental breakthroughs in manipulating ultracold atoms in optical lattices have opened up the possibility to perform quantum simulations of the Fermi-Hubbard model \cite{2010Esslinger,Boll2016,Cheuk2016b,Mazurenko2017,Hilker2017,Brown2017,Chiu2018,Brown2019,Koepsell2019,Chiu2019,Koepsell2020,Brown2020a,Vijayan2020,Hartke2020,Brown2020b,Ji2021,Gall2021}. In particular, the ability to image individual atoms with single-site spatial resolution \cite{Bakr2009,Sherson2010,Haller2015,Yang2021} makes it possible to probe the microscopic structure of the magnetic polaron \cite{Koepsell2019,Chiu2019}. Such detailed insights offer stringent tests of the understanding of the Fermi-Hubbard model, and enable a systematic improvement of theoretical approaches to these quasiparticles.

A particularly successful approach has been the self-consistent Born approximation (SCBA) \cite{SchmittRink1988,Kane1989}. The SCBA permits a non-perturbative calculation of the Green's function of the hole and was shown to yield quantitatively accurate results \cite{Martinez1991,Liu1991,Liu1992,Chernyshev1999} in the strongly interacting regime. Despite this success, its applicability has thus far been limited to single-particle observables such as the energy dispersion and quasiparticle residue of the magnetic polaron, while the extraction of finer structural information, such as that contained in spin-hole correlations or the polaron's magnetic dressing cloud, have proved difficult. The construction of the polaron wave function in terms of spin-wave excitations \cite{Reiter1994} in principle offers a solution to this problem. Determining correlation functions, however, entails an infinite series of terms with increasing number of spin excitations \cite{Ramsak1993,Ramsak1998}, whose truncation has restricted calculations to the weak coupling domain \cite{Ramsak1993,Ramsak1998,Bala2000,Bala2002}. Moreover, these conditions violate the underlying assumptions of the SCBA treatment \cite{SchmittRink1988,Kane1989} in the context of the Fermi-Hubbard model.

Here, we develop a theoretical framework that makes it possible to overcome this obstacle. The approach is based on a set of self-consistency equations that are reminiscent of the Dyson equation, and enable the inclusion of \emph{all} terms up to infinite numbers of spin excitations in the SCBA wave function for the magnetic polaron. It thereby extends its application into the regime of strong interactions. We use this new approach to explore the microscopic spatial structure of the magnetic polaron in the regime of strong coupling. In general, the obtained magnetic dressing cloud has an elongated shape that increases in size and magnitude with the strength of interactions, and, in the strong-coupling regime, differs qualitatively from previous calculations based on a truncated quasiparticle wave function. Our analysis reveals that the symmetries of the antiferromagnetic spin lattice decisively determine the form of the dressing cloud, and lead to a remarkably high symmetry of the magnetic dressing cloud for polaron momenta along the edge of the magnetic Brillouin zone (MBZ). Surprisingly and in contrast to previous expectation, this can lead to a misalignment between the dressing cloud and the crystal momentum, such that the spatial structure of the polaron is generally not oriented along its direction of motion.

Our theoretical framework moreover permits to explore the transition of the underlying quantum magnet from the isotropic Heisenberg spin-lattice to the Ising model. We find that the gap opening in the spin-wave excitation spectrum results in a shrinking of the polaronic magnetization cloud with increasing anisotropy of the effective spin-spin interaction. In the Ising limit, the full symmetry of the antiferromagnetic order is restored, which makes it possible to determine the dressing cloud of the magnetic polaron analytically. More generally, the developed framework may open up a new approach for microscopic explorations of doped quantum magnets in the strong-coupling regime, including finite temperature effects and non-equilibrium dynamics to induced interactions between multiple defects. 

This article is organised as follows. Section \ref{sec.t_J_model} provides the Fermi-Hubbard Hamiltonian and the $t$-$J$ model that derives from it. Based on the $t$-$J$ model, we describe the transformation into a magnetic polaron Hamiltonian within linear spin wave theory. In Sec. \ref{sec.magnetic_polaron_quasiparticle}, we summarize the quasiparticle properties of the magnetic polaron, including the calculation of the polaron Green's and wave functions within the SCBA. The magnetization in the vicinity of a hole is explored in Sec.\ \ref{sec.local_magnetization} for a two-dimensional square lattice as a function of interaction strength and anisotropy. In Sec.\ \ref{sec.non_perturbative_effects}, we demonstrate the non-perturbative effects predicted by our developed formalism, and Sec.\ \ref{sec.derivation} gives a detailed derivation of the self-consistency equations for the local magnetization. Finally, in Sec.\ \ref{sec.experiments}, we describe the prospects of testing our theory experimentally. 

\section{The anisotropic $t$-$J$ model} \label{sec.t_J_model}
The Fermi-Hubbard model 
\begin{equation}
\hat{H}_{\rm FH} = -t \sum_{\braket{{\bf i}, {\bf j}}, \sigma} \left[ \hat{c}^\dagger_{{\bf i},\sigma} \hat{c}_{{\bf j},\sigma} + {\rm H.c.} \right] + U \sum_{\bf i} \hat{n}_{{\bf i},\uparrow}\hat{n}_{{\bf i},\downarrow}
\label{eq.H_FH}
\end{equation}
describes spin-1/2 fermions moving in a lattice with hopping amplitude $t$ and onsite repulsive interactions $U > 0$. Here, $\braket{{\bf i}, {\bf j}}$ denotes nearest neighbor lattice sites, $\hat{c}^\dagger_{{\bf i},\sigma}$ creates a fermion at site $\bf i$ and spin $\sigma$, while $\hat{n}_{{\bf i},\sigma} = \hat{c}^\dagger_{{\bf i},\sigma}\hat{c}_{{\bf i},\sigma}$ is the corresponding counting operator. Despite its apparent simplicity, many open questions remain concerning its properties. Near half filling, one can expand the Hubbard model in the particle hopping for large onsite repulsion $U \gg t$ to derive an effective low-energy description given by the so-called $t$-$J$ model \cite{Chao1977,Dagotto1994,Izyumov1997}. The Hamiltonian is $\hat{H} = \hat{H}_t + \hat{H}_J$, where
\begin{equation}
\hat{H}_t = -t \sum_{\braket{{\bf i}, {\bf j}}, \sigma} \left[ \tilde{c}^\dagger_{{\bf i},\sigma} \tilde{c}_{{\bf i},\sigma} + {\rm H.c.} \right], 
\label{eq.H_t}
\end{equation}
describes the restrained nearest neighbor particle hopping, where $\tilde{c}^\dagger_{{\bf j},\sigma}=\hat{c}_{{\bf j},\sigma}^\dagger(1 - \hat n_{{\bf j}, \bar{\sigma}})$, and the factor $(1 -\hat n_{{\bf j}, \bar{\sigma}})$ with 
$\bar{\sigma}$ the opposite spin restrains the Hilbert space of the model to states with maximally one particle per site. Furthermore,
\begin{equation}
\hat{H}_J = J \sum_{\braket{{\bf i}, {\bf j}}} \!\left[ \hat{S}^{z}_{\bf i} \hat{S}^{z}_{\bf j} \!+\! \frac{\alpha}{2}\!\left(\hat{S}^{+}_{\bf i} \hat{S}^{-}_{\bf j} \!+\! \hat{S}^{-}_{\bf i} \hat{S}^{+}_{\bf j}\right) \!-\! \frac{\hat{n}_{\bf i} \hat{n}_{\bf j}}{4} \right]
\label{eq.H_J}
\end{equation}
gives the antiferromagnetic ($J > 0$) spin-spin interactions. The Schwinger-fermion representation of spin $1/2$ reads as usual 
\begin{equation}
\mathbf{S}_{\bf j}=\frac{1}{2}\sum_{\sigma,\sigma'}\hat{c}_{{\bf j},\sigma}^\dagger\boldsymbol{\sigma}_{\sigma\sigma'}\hat{c}_{{\bf j},\sigma'}
\end{equation}
with $\boldsymbol{\sigma}=(\sigma_x,\sigma_y,\sigma_z)$ a vector of the Pauli matrices. The $t$-$J$ model with $\alpha = 1$ thus yields an accurate description of the low-energy physics of the underlying Fermi-Hubbard model for $t \gg J = 4t^2 / U$ close to half filling. More generally, other experimental platforms \cite{Porras2004,Gorshkov2011_2,Britton2012,Zeiher2016,Zeiher2017} make it possible to tune from the isotropic Heisenberg limit with $\alpha=1$ to an Ising magnet with $\alpha=0$. In Eq. \eqref{eq.H_J}, we have therefore generalized the model to include the case of anisotropic spin interactions by introducing the parameter $0 \leq \alpha \leq 1$.

At half filling with exactly one fermion per lattice site, the first term in Eq. \eqref{eq.H_t} is ineffective and a positive superexchange coupling $J > 0$ between the spins enforces antiferromagnetic ordering for any value of $\alpha$. Lattice defects, or holes, in such an antiferromagnet tend to delocalize and thereby lower their kinetic energy, as given by $\hat{H}_t$. The associated motion of holes, on the other hand, leads to the buildup of magnetic frustration which increases the energy of the system according to $\hat{H}_J$. The competition between these two processes eventually gives rise to the magnetic polaron, i.e. a mobile hole that is surrounded by a finite magnetization cloud. Small ratios of $J / t$, thus, correspond to the strong coupling regime in which a high hole mobility leads to a significant disturbance of its magnetic environment and thereby generates strong spin-hole correlations. We can accurately describe this process using spin-wave theory as outlined in the next section.

\subsection{Slave fermion representation}
We begin by performing a Holstein-Primakoff transformation generalized to take into account the presence of holes \cite{SchmittRink1988,Kane1989,Martinez1991,Liu1991}. The antiferromagnetic state defines a bipartite lattice, whereby one sublattice carries fermions in the spin-up state, while the other sublattice is formed by particles in the spin-down state. In the former, we rewrite $\hat{S}_{\bf i}^{-}=\hat{s}_{\bf i}^\dagger(1-\hat{s}_{\bf i}^\dagger\hat{s}_{\bf i}-\hat h^\dagger_{\bf i}\hat h_{\bf i})^{1/2}$, $\tilde{c}_{{\bf i},\downarrow}=\hat{h}^\dagger_{\bf i}\hat{s}_{\bf i}$, and $\tilde{c}_{{\bf i},\uparrow}=\hat h^\dagger_{\bf i}(1-\hat s^\dagger_{\bf i}\hat{s}_{\bf i}-\hat{h}^\dagger_{\bf i}\hat{h}_{\bf i})^{1/2}$ in terms of fermionic operators $\hat{h}^\dagger_{\bf i}$ and bosonic operators $\hat{s}^\dagger_{\bf i}$ that create a hole and a spin excitation at site ${\bf i}$ respectively. The factor $(1-\hat{s}^\dagger_{\bf i}\hat{s}_{\bf i}-\hat{h}^\dagger_{\bf i}\hat{h}_{\bf i})^{1/2}$ ensures that there is at most one hole or one spin excitation at each site. Finally, the spin-$z$ operator can be rewritten as $\hat{S}_{\bf i}^z=(1-\hat{h}_{\bf i}^\dagger \hat{h}_{\bf i})/2-\hat{s}^\dagger_{\bf i}\hat{s}_{\bf i}$. The representation of the spin and holes on the other sublattice of spin-down fermions, proceeds analogously by swapping spin $\uparrow$ and $\downarrow$ in the transformations given above. Using this so-called slave-fermion representation in Eq. \eqref{eq.H_J}, keeping only the linear terms, and diagonalizing the transformed Hamiltonian yields the spin wave Hamiltonian \cite{SchmittRink1988,Kane1989,Martinez1991,Liu1991}
\begin{align}
H_J = E_0 + \sum_\bk \omega_\bk \hat{b}^\dagger_\bk \hat{b}_\bk.
\label{eq.H_J_diagonalized}
\end{align}
This describes spin waves with the energy $\omega_\bk = zJ \sqrt{1 - \alpha^2 \gamma_\bk^2} / 2$, where the structure factor 
\begin{align}
\gamma_\bk = \frac{1}{z}\sum_{\bdelta} \te^{i\bk\cdot \bdelta},
\label{eq.gamma_k}
\end{align}
is the sum of the $z$ nearest neighbor phases. The associated bosonic spin wave operators $\hat{b}_\bk$ are related to the physical spin excitations $\hat s_\bk$ via a Bogoliubov transformation $\hat{b}_\bk = u_\bk \hat{s}_\bk + v_\bk \hat{s}^\dagger_{-\bk}$ with the antiferromagnetic coherence factors given by $u_\bk = [(1 / \sqrt{1 - \alpha^2 \gamma_\bk^2} + 1 ) / 2]^{1/2}$ and $v_\bk = {\rm sgn}(\gamma_\bk) [(1 / \sqrt{1 - \alpha^2 \gamma_\bk^2} - 1 ) / 2]^{1/2}$. Using the slave fermion representation in Eq. \eqref{eq.H_t} yields \cite{SchmittRink1988,Kane1989,Martinez1991,Liu1991}
\begin{align}
\hat{H}_t = \sum_{\bq, \bk} \hat{h}_{\bq + \bk}^\dagger \hat{h}_{\bq} \left[ g(\bq, \bk) \hat{b}^\dagger_{-\bk} + g(\bq + \bk, -\bk) \hat{b}_\bk \right],
\label{eq.H_t_k}
\end{align}
where the interaction vertex is $g(\bq, \bk) = zt \cdot (u_\bk \gamma_{\bq + \bk} - v_\bk \gamma_\bq) / \sqrt{N}$ with $N$ the number of lattice sites. Here, we retain terms linear in the spin-wave operators. Equation \eqref{eq.H_t_k} 
explicitly shows how the hopping of a hole gives rise to the emission/absorption of spin waves and directly represents the competition between hole delocalization and magnetic order.

\section{Magnetic polaron} \label{sec.magnetic_polaron_quasiparticle}
The above formulation in terms of interactions with magnetic spin-wave excitations enables a non-perturbative description of the microscopic structure of the magnetic polaron, as we shall outline in this section. The approach is based on the self-consistent Born approximation (SCBA) for the hole Green's function \cite{SchmittRink1988,Kane1989}, which has been shown to yield quantitatively accurate results for the polaron energy in the Heisenberg limit across all interaction strengths \cite{Martinez1991}. 

\subsection{The Green's function}
\begin{figure}[t!]
\center
\begin{center}
\includegraphics[width=\columnwidth]{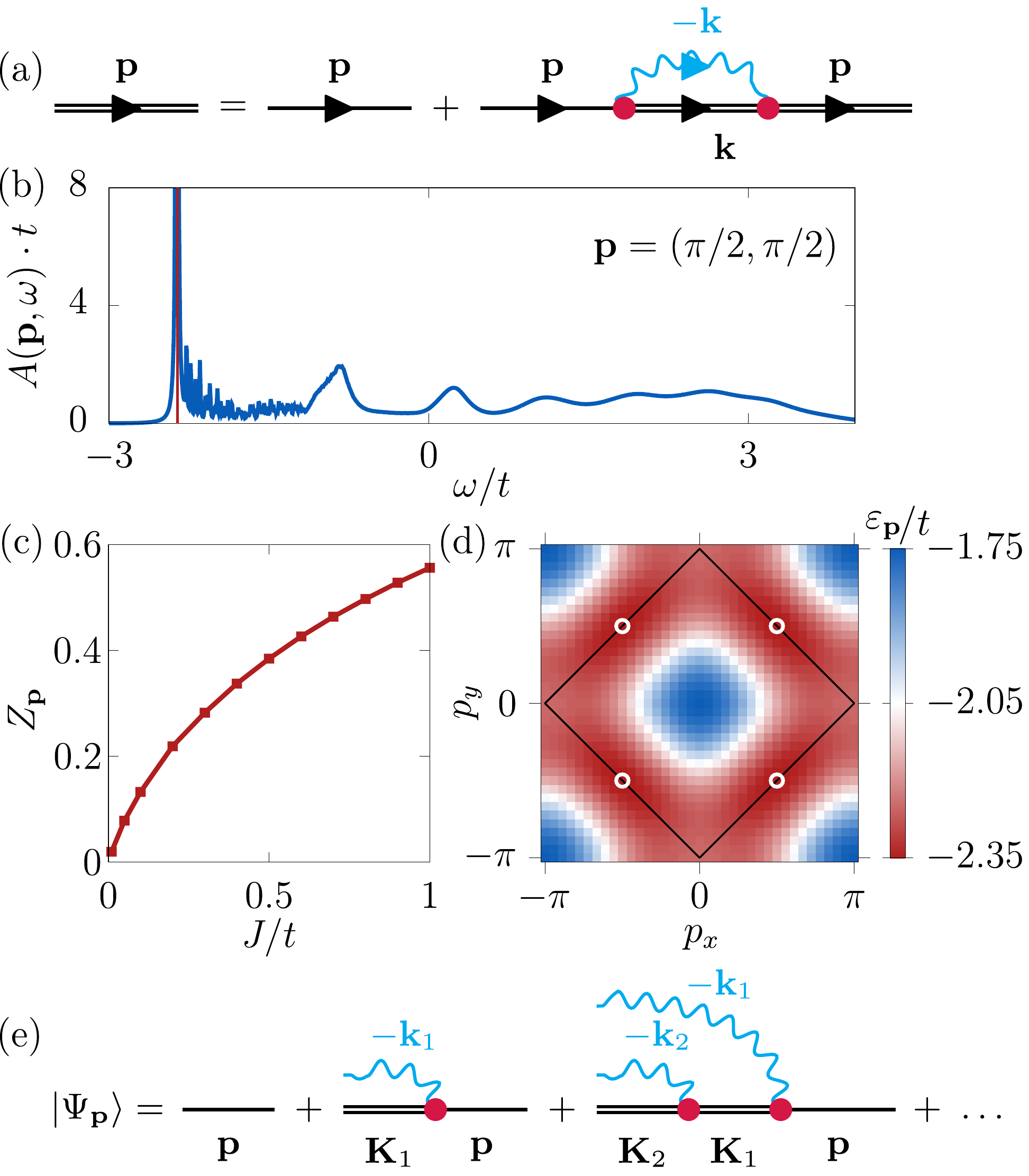}
\end{center}
\caption{\textbf{Quasiparticle properties.} (a) Dyson equation for the hole Green's function, $G(\bp, \omega)$, within SCBA (double black line), in terms of the non-interacting Green's function $G_0(\omega) = 1 / \omega$ (single black line), and the spin wave Green's function $G_{b} = 1 / (\omega - \omega_\bk)$ (blue wiggly line). (b) Hole spectral function $A(\bp, \omega) = -2{\rm Im} G(\bp, \omega)$ in the Heisenberg limit in a 36 by 36 square lattice, featuring a quasiparticle peak at $\varepsilon_\bp \simeq -2.4t$ for $J / t = 0.3$ (vertical red line). For $J / t \ll 1$, the quasiparticle residue [panel (c)] becomes very small. Consequently, an increasing number of terms must be retained in the quasiparticle wave function [panel (e)]. The diagrammatic rules for the construction of the wave function are: (1) Single straight line: $Z_\bp^{1/2}$. (2) Blue wiggly line with momentum $-\bk_i$: spin wave operator, $\hat{b}^\dagger_{-\bk_i}$. (3) nth red dot from the right in a diagram: interaction vertex, $g(\bK_n, \bk_{n+1})$. Here, $\bK_0 = \bp$, $\bK_1 = \bp + \bk_1$, $\bK_2 = \bp + \bk_1 + \bk_2$, and so forth. (4) nth double line from the right: $G\left(\bK_n, \varepsilon_\bp - \sum_{i = 1}^n \omega_{\bk_i}\right)$. (5) nth open-ended hole line from the right: hole operator, $\hat{h}^\dagger_{\bK_n}$. (6) Sum over all spin wave momenta, $\bk_i$. (d) The quasiparticle dispersion throughout the first Brillouin zone with the four degenerate ground states at $\bp = (\pm\pi/2, \pm\pi/2)$ indicated by circles. The magnetic Brillouin zone (MBZ) is indicated by black diagonal lines.}
\label{fig.quasiparticle_properties}
\end{figure}

The SCBA includes the so-called rainbow diagrams in the computation of the hole Green's function $G(\bp, \omega) = 1 / (\omega - \Sigma(\bp, \omega) + i\eta)$ \cite{SchmittRink1988,Kane1989,Martinez1991,Liu1991}, where $\eta = 0^+$ is a positive infinitesimal. Using the spin wave Green's function $G_b(\bk, \omega) = 1 / (\omega - \omega_\bk + i\eta)$, the diagrammatic structure shown in Fig. \ref{fig.quasiparticle_properties}(a) leads to the self-consistent equation for the self-energy
\begin{equation}
\Sigma(\bp, \omega) = \sum_\bk \frac{g^2(\bp, \bk)}{\omega - \omega_\bk - \Sigma(\bp + \bk,\omega - \omega_\bk) + i\eta},
\label{eq.hole_self_energy_self_consistency_equation}
\end{equation}
which can be solved iteratively starting from $\Sigma = 0$. Knowing the self-energy, one can determine several important quantities, such as the quasiparticle residue 
\begin{align}
Z_\bp = \frac{1}{1 - \pa_\omega \Sigma(\bp, \omega)|_{\omega = \epsilon_\bp}},
\label{eq.quasiparticle_residue}
\end{align}
which is the overlap $Z_\bp=|\bra{\AF}\hat h_{\bp}\ket{\Psi_\bp}|^2$ of the polaron many-body wave function $\ket{\Psi_\bp}$ with the state of a bare hole in an otherwise unperturbed antiferromagnetic state, $\hat h_{\bp}^\dagger\ket{\AF}$, for a given crystal momentum $\bp$ of the hole. Hereby, the antiferromagnetic quantum N{\'e}el state is defined as $\hat{b}_{\bk}\ket{\AF} = 0$ for any spin-wave momentum $\bk$. The numerical solution of Eq. \eqref{eq.hole_self_energy_self_consistency_equation} under strong-coupling conditions, $J/t=0.3$, in a 36 by 36 square lattice is shown in Fig. \ref{fig.quasiparticle_properties}. The hole spectral function, depicted in Fig. \ref{fig.quasiparticle_properties}(b), exhibits a clear quasiparticle peak at $\omega\simeq-2.4$t giving the energy of the magnetic polaron. The corresponding quasiparticle residue is $Z \simeq 0.3$ and there is a continuum of many-body states at higher energies with large spectral weight, reflecting the strongly interacting nature of the problem. The dispersion of the magnetic polaron in the first Brillouin zone is shown in Fig. \ref{fig.quasiparticle_properties}(d). It features four degenerate ground states at the crystal momenta $\bp = (\pm \pi / 2, \pm \pi / 2)$, given in units of the inverse lattice constant. The surprising predictive power of the SCBA result for the hole spectral function in the Heisenberg limit \cite{Martinez1991} compared to exact diagonalization studies \cite{Dagotto1990} is attributed to small vertex corrections to the SCBA even for strong coupling \cite{Liu1992}. The dependence of quasiparticle residue on $J/t$ [Fig. \ref{fig.quasiparticle_properties}(c)] clearly illustrates the necessity of a non-perturbative theory for $J / t \ll 1$, since small values of $Z_{\bp}$ indicate a large number of spin-wave excitations and strong correlations between the generated spin fluctuations and the motion of the hole. The Green's function alone is, however, not well suited to study such correlations and requires additional analysis as we will now discuss.

\subsection{The polaron wave function}
Our non-perturbative approach to determine spin-hole correlations exploits the fact that it is formally possible to write the wave function $\ket{\Psi_\bp}$ of the magnetic polaron within the SCBA. Explicitly, the wave function \cite{Reiter1994,Ramsak1998}
\begin{align}
&\ket{\Psi_\bp} = \sum_{n = 0}^\infty \sum_{\{\bk_i\}} a^{(n)}(\bp, \{\bk_i\}_{i=1}^n) \cdot \hat{h}^\dagger_{\bK_n}\prod_{i=1}^n \hat{b}_{-\bk_i}^\dagger\ket{\AF} \nn \\
&= \sqrt{Z_\bp} \Big[h^\dagger_\bp \!+\! \sum_{\bk_1} g(\bp, \bk_1) G\left(\bp \!+\! \bk_1, \varepsilon_\bp \!-\! \omega_{\bk_1}\right) \hat{h}^\dagger_{\bp + \bk_1} \hat{b}^\dagger_{-\bk_1} \nn \\
& \phantom{= \sqrt{Z_\bp} \Big[} + \dots \Big] \ket{\AF},
\label{eq.Reiters_wave_function}
\end{align}
can be expressed as an expansion in the number of spin wave excitations on the antiferromagnetic quantum N{\'e}el state $\ket{\AF}$. The lowest order coefficient, $a^{(0)}(\bp) = \sqrt{Z_\bp}$, is given by the square root of the quasiparticle residue ensuring the overall normalization of the wave function, while the higher order coefficients can be computed from the recurrence relation 
\begin{align}
a^{(n+1)}(\bp, \{\bk_i\}_{i=1}^{n+1}) =& g(\bK_n, \bk_{n+1}) a^{(n)}(\bp, \{\bk_i\}_{i=1}^{n})\nn\\
&G\left(\bK_{n + 1}, \varepsilon_\bp - \sum_{i=1}^n \omega_{\bk_i} \right), 
\label{eq.recursion_relation}
\end{align}
with $\bK_n = \bp + \sum_{i = 1}^n \bk_i$ for $n \geq 1$, and $\bK_0 = \bp$. The structure of the first order term $a^{(1)}$ can be understood by using the recursion relation in Eq. \eqref{eq.recursion_relation} in the coupling between the two lowest order coefficients, $a^{(0)}$ and $a^{(1)}$, of the quasiparticle wave function \eqref{eq.Reiters_wave_function}. This yields $\varepsilon_{\bf p} a^{(0)}(\bp) = \sum_{\bk} g(\bp, \bk) a^{(1)}(\bp,\bk) = \Sigma(\bp, \varepsilon_\bp) a^{(0)}(\bp)$, by using Eq. \eqref{eq.hole_self_energy_self_consistency_equation} for the self-energy $\Sigma$. Hence, the construction of the wave function \eqref{eq.Reiters_wave_function} relies on the presence of a well-defined quasiparticle peak at $\varepsilon_\bp$ determined by $\varepsilon_\bp = \Sigma(\bp, \varepsilon_\bp)$, corresponding to the energy of the magnetic polaron.  

The wave function is visualized diagrammatically \cite{Ramsak1998} in Fig. \ref{fig.quasiparticle_properties}(e), along with the diagrammatic rules for its construction. The iterative structure of Fig. \ref{fig.quasiparticle_properties}(e) together with precise diagrammatic rules is reminiscent of the Dyson equation in quantum field theory [Fig. \ref{fig.quasiparticle_properties}(a)], and is at the heart of our non-perturbative framework developed below. A major advantage of the polaron wave function is that it allows for the computation of spin-hole correlation functions in a much more direct way than the hole Green's function. The problem is nevertheless still far from straightforward, since the number of important terms in Eq. \eqref{eq.Reiters_wave_function} increases with the interaction, $t/J$. As a consequence, there is no controlled way to truncate the series for the wave function, while still obtaining reliable results in the strong coupling regime $J / t \ll 1$. This has so far limited the use of this wave function to the weak coupling regime, where it is sufficient to include only a small of number spin wave excitations \cite{Ramsak1993,Ramsak1998}. However, since the SCBA is least accurate precisely in the weak coupling regime of $J / t \gtrsim 1$, the validity of such an approach is not clear. The non-perturbative framework, developed in the present work, thus represents a major step as it now makes it possible to utilize the SCBA for reliable calculations of correlation functions in the strongly coupled regime.

\section{Magnetization around a hole} \label{sec.local_magnetization} 
Before presenting the derivation of our non-perturbative approach in Sec.\ \ref{sec.derivation}, in this section, we first illustrate its application by calculating the local magnetization in the neighborhood of a hole. This is a fundamental property of the magnetic polaron that determines its microscopic structure, and eventually the form of the induced interaction between multiple holes. 

Consider first the magnetization in the absence of holes. The local magnetization at a given lattice site $\bd$ is given by
\begin{equation}
\!\!2\bra{\AF}\!\hat{S}^{z}_\bd\!\ket{\AF} = (-1)^{l_\bd}M_\text{AF} = (-1)^{l_\bd}\!\left(1 \!-\! 2M_\text{fl}\right),\!\!
\label{eq.spin_magnetization_no_hole}
\end{equation}
where $0 \leq M_\text{fl} \leq 1$ quantifies the effect of quantum spin fluctuations to suppress the magnetic order from its maximum value $M_\text{AF} = 1$ in the Ising limit $\alpha=0$. Without loss of generality, we take $\bra{\AF}\hat{S}^{z}_{\bd = {\bf 0}}\ket{\AF} > 0$ in the spontaneously broken symmetry state $\ket{\AF}$. The site-dependent parameter $l_\bd$ is defined as the minimal number of lattice links between the two sites $\br = \mathbf 0$ and $\br = \bd$. From linear spin-wave theory, we have $\hat{S}^{z}_{\bd} = (-1)^{l_\bd} (1 / 2 - \hat{s}^\dagger_{\bd} \hat{s}_{\bd})$ and $\hat{s}^\dagger_{\bd} = \sum_\bk \te^{-i\bk\cdot\bd}\hat{s}_{\bk}^\dagger / \sqrt{N}$, which gives $M_\text{fl}= \sum_\bk\bra{\AF}\hat{s}_{\bk}^\dagger\hat{s}_{\bk}\ket{\AF}/N=\sum_{\bk} v_\bk^2 /N$. 

With these definitions, we can now formulate the magnetization of a given lattice site at a distance $\bd$ from the hole 
\begin{align}
M_\bp(\bd) = \frac{\braket{\hat{h}^\dagger_{\br}\hat{h}_{\br} \hat{S}^{z}_{\br+\bd}}_\bp}{\braket{\hat{h}^\dagger_{\br}\hat{h}_{\br}}_\bp\braket{\hat{S}^{z}_{\br + \bd}}_\bp} = \frac{1 - 2 M_\bp^{(2)}(\bd)}{M_\text{AF}},
\label{eq.spin_magnetization_1}
\end{align}
where $\braket{\ldots}_\bp = \bra{\Psi_\bp} \ldots \ket{\Psi_\bp}$ is the expectation value for the ground state Eq. \eqref{eq.Reiters_wave_function} of a magnetic polaron with crystal momentum $\bp$, and 
\begin{align}
M_\bp^{(2)}(\bd) = N\braket{\hat{h}^\dagger_{\br}\hat{h}_{\br}\hat{s}^\dagger_{\br + \bd} \hat{s}_{\br + \bd}}_\bp.
\label{eq.M_2}
\end{align}
The translational symmetry of the system ensures that these correlation functions only depend on the distance vector $\bd$. In Eqs. \eqref{eq.spin_magnetization_1} and \eqref{eq.M_2}, we have used $\braket{\hat{h}^\dagger_{\br}\hat{h}_{\br}}_\bp = 1 / N$, reflecting the fact that the hole is equally distributed across the lattice for a given momentum state. We can also omit corrections to the average magnetization from the presence of a single hole, since they scale as $\braket{\hat{S}^{z}_{\bd}}_\bp-\bra{\AF}\hat{S}^{z}_{\bd}\ket{\AF}\sim{\mathcal O}(1/N)$. 

Fourier transforming and rotating to the bosonic spin wave operators $\hat{b}_\bk$, the remaining two-point correlator $M_\bp^{(2)}$ given by Eq. \eqref{eq.M_2} can be decomposed as
\begin{align}
M^{(2)}_\bp(\bd) = M_\text{fl} + B_\bp(\bd) + C_\bp(\bd).
\label{eq.spin_magnetization_2}
\end{align}
Notice that the zero-point fluctuations $M_\text{fl}$ of the quantum antiferromagnet appears explicitly. The corrections to the magnetization due to the presence of the hole are thus 
described by the functions
\begin{align}
&\!\!\!\!B_\bp(\bd) = \frac{1}{N^2} \!\!\sum_{\bq_1, \bq_2} \te^{-i(\bq_2 - \bq_1)\cdot \bd} \left[u_{\bq_1} u_{\bq_2} + v_{\bq_1}v_{\bq_2}\right] \nn \\
&\phantom{\!\!\!\!B_\bp(\bd) = \frac{1}{N^2}\!\! \sum_{\bq_1, \bq_2}} \cdot B(\bq_1, \bq_2; \bp, \varepsilon_\bp), \nn \\
&\!\!\!\!B(\bq_1, \bq_2; \bp, \omega) = N \sum_\bk \braket{\hat{h}^\dagger_{\bk + \bq_1} \hat{h}_{\bk + \bq_2} \hat{b}^\dagger_{-\bq_1} \hat{b}_{-\bq_2}}_{\bp, \omega},
\label{eq.B_position_space}
\end{align}
and
\begin{align}
&\!\!\!\!C_\bp(\bd) = -\frac{1}{2 N^2}\!\!\sum_{\bq_1, \bq_2} \te^{-i(\bq_2 + \bq_1)\cdot \bd} \left[u_{\bq_1} v_{\bq_2} + v_{\bq_1}u_{\bq_2}\right] \nn \\
&\phantom{\!\!\!\!C(\bd)= -\frac{1}{N^2}\!\! \sum_{\bq_1, \bq_2}} \cdot C(\bq_1, \bq_2; \bp, \epsilon_\bp) + \textrm{c.c.} \nn \\
&\!\!\!\!C(\bq_1, \bq_2; \bp, \omega) = N \sum_{\bk} \braket{\hat{h}^\dagger_{\bk - \bq_1 - \bq_2}\hat{h}_\bk \hat{b}_{-\bq_1} \hat{b}_{-\bq_2}}_{\bp, \omega},
\label{eq.C_position_space}
\end{align}
where c.c. stands for the complex conjugate. In the definition of $B_\bp(\cdot, \omega)$ and $C_\bp(\cdot, \omega)$, we allow the energy $\omega$ of the magnetic polaron to vary, meaning that the expectation value $\braket{\ldots}_{\bp,\omega}$ is taken with respect to the state $\ket{\Psi_\bp}$ given by Eq. \eqref{eq.Reiters_wave_function}, 
where the polaron energy $\varepsilon_\bp$ is replaced by $\omega$ in the appearing Green's functions. As we will describe in Sec.\ \ref{sec.derivation}, this generalization combined with the diagrammatic rules for the wave function given in Fig. \ref{fig.quasiparticle_properties} makes it possible to derive self-consistency equations for the correlation functions in Eqs. \eqref{eq.B_position_space} and \eqref{eq.C_position_space}. These equations are similar to Eq. \eqref{eq.hole_self_energy_self_consistency_equation} for the self-energy, and ultimately facilitate evaluation of correlation functions to \emph{all} orders in the number of spin-wave excitations in the wave function Eq. \eqref{eq.Reiters_wave_function}. 

\subsection{Heisenberg limit} \label{subsec.heisenberg_limit}
In Figs. \ref{fig.magnetization_fig1}(a) and \ref{fig.magnetization_fig1}(b), we show the spatial structure of the local magnetization $\tilde{M}_\bp(\bd) = 2\braket{\hat{S}^{z}_{\bd}}_\bp M_\bp(\bd)$ in the isotropic Heisenberg limit, $\alpha \to 1$, for two different coupling strengths. The crystal momentum is $\bp = (\pi/2,\pi/2)$, which corresponds to one of the four degenerate ground states of the polaron [Fig. \ref{fig.quasiparticle_properties}(d)]. The results show that the hole significantly reduces the magnetic order in its vicinity. This effect is particularly prominent for strong coupling, $J / t = 0.05$, where the size of the magnetization cloud is increased, and where the nearest neighbor spins are even flipped as a consequence of strong correlations between the motion of the hole and the local magnetization of the lattice. 
\begin{figure}[t]
\begin{center}
\includegraphics[width=\columnwidth]{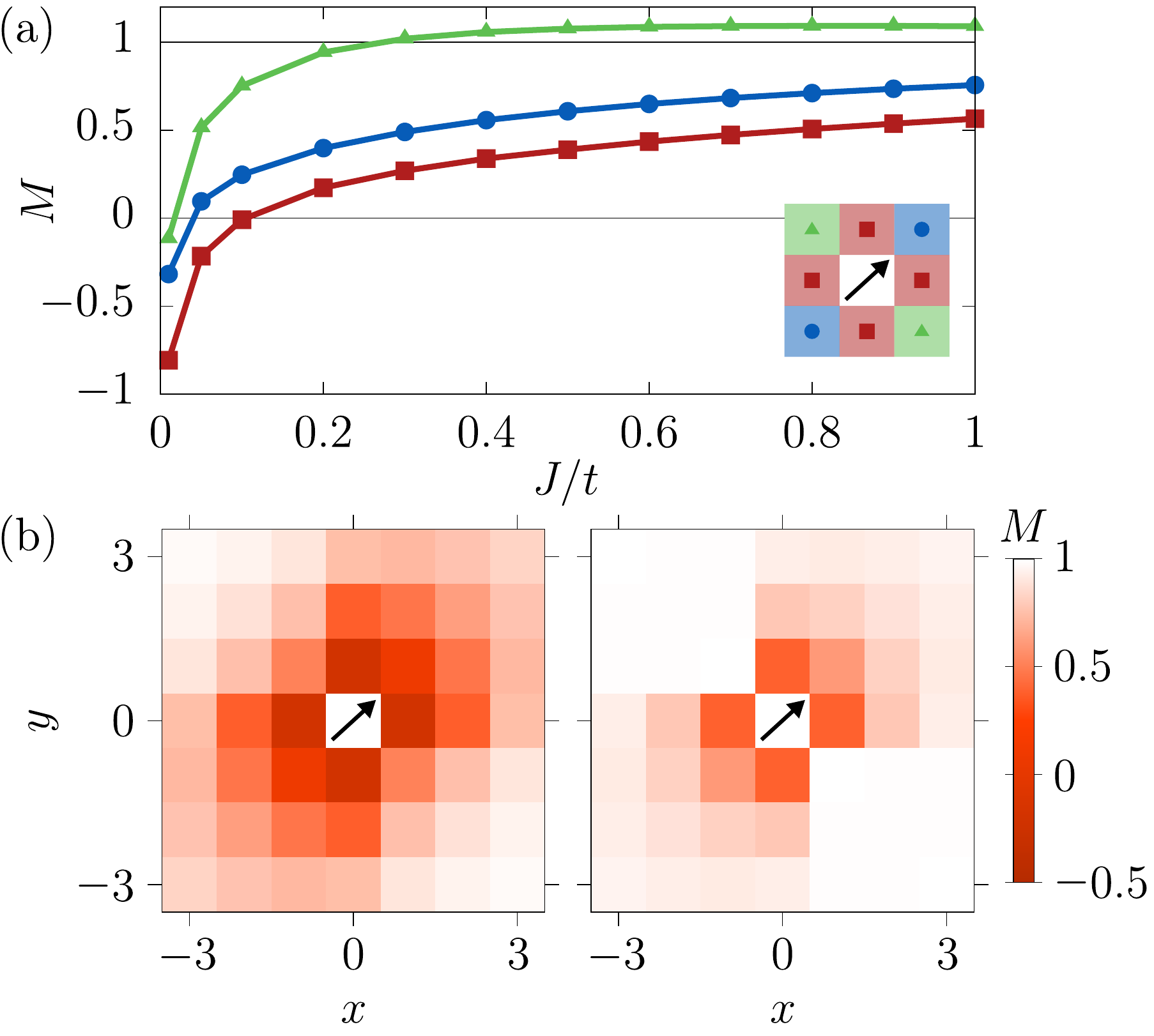}
\end{center}
\caption{Magnetization in the neighborhood of a hole for the ground state magnetic polaron with $\bp=(\pi/2,\pi/2)$ in a Heisenberg antiferromagnet as a function of interaction strength, $J / t$, in a 16 by 16 square lattice. (a) Red squares are for the nearest neighbor sites, blue dots are the next nearest neighbors, and green triangles are for the perpendicular direction as illustrated in the inset. The lines are guides to the eye. Note that the magnetization at the green points are larger than in the absence of holes for intermediate interaction strengths, $J / t > 0.3$. (b) Heat maps of the magnetization for a strong and intermediate interaction strength, $J / t = 0.05$ (left) and $J / t = 0.5$ (right) respectively.}
\label{fig.magnetization_J_vary} 
\end{figure} 

The emergence of this sign flip is illustrated in Fig.\ \ref{fig.magnetization_J_vary}(a), where we show the magnetization $M_{\bp}(\bd)$ as a function of the inverse interaction strength $J/t$, indicating that this strong magnetic disturbance extends to larger and larger distances as we enter the strong coupling regime and further increase $t/J$. The results also show that the magnetization of the next-nearest neighbors is anisotropic and reflects the direction of the crystal momentum of the moving polaron. Perpendicular to this direction, the magnetization can even be larger than in the absence of the hole. This surprising effect results from the coherent addition of the generated spin waves to produce a net increase in the magnetization for intermediate interaction strengths. 

The elongated shape of the magnetization cloud is shown more directly in Fig. \ref{fig.magnetization_J_vary}(b). It appears that the magnetization cloud is oriented along the crystal momentum of the polaron, as also reported previously based on truncated wave function calculations. There, the observed alignment has been attributed to the semiclassical idea that the hole will predominantly disturb the magnetization in the direction of its motion \cite{Ramsak1993,Ramsak1998}. However, as we will discuss below, this is \emph{generally not the case}, as the symmetry properties of the underlying antiferromagnetic competes with the directed motion of the polaron and yields a nontrivial orientation of the magnetization cloud with respect to the polaron crystal momentum. 

While the SCBA can in principle lead to an unphysical spin state in the vicinity of the hole, we find no evidence for this in the entire investigated region. More precisely, the physical limits of the magnetization in Eq. \eqref{eq.spin_magnetization_1} is $\pm 1 / (1 - 2M_{\rm fl}) \simeq \pm 1.67$ corresponding to having exactly $0$ or $1$ spin excitations at a given site. The most extreme value is associated with the nearest neighbor magnetization $M(d = 1) \simeq -0.81$ at $J / t = 0.01$. This corresponds to a mean value of spin excitations of $M^{(2)}(d = 1) = 0.74$.

\subsection{Symmetries} \label{subsec.spatial_symmetries}
The preceding discussion suggests an underlying symmetry of the magnetization around the hole, which we will explore for general crystal momentum of the polaron in this section. The antiferromagnetic spin lattice exhibits several point symmetries, namely mirror symmetries with respect to the two principal axes and the diagonals of the lattice, as well as $C_4$ rotations. This combines to the symmetry group $C_{4v}$. 

\begin{figure*}[t]
\begin{center}
\includegraphics[width=1.35\columnwidth]{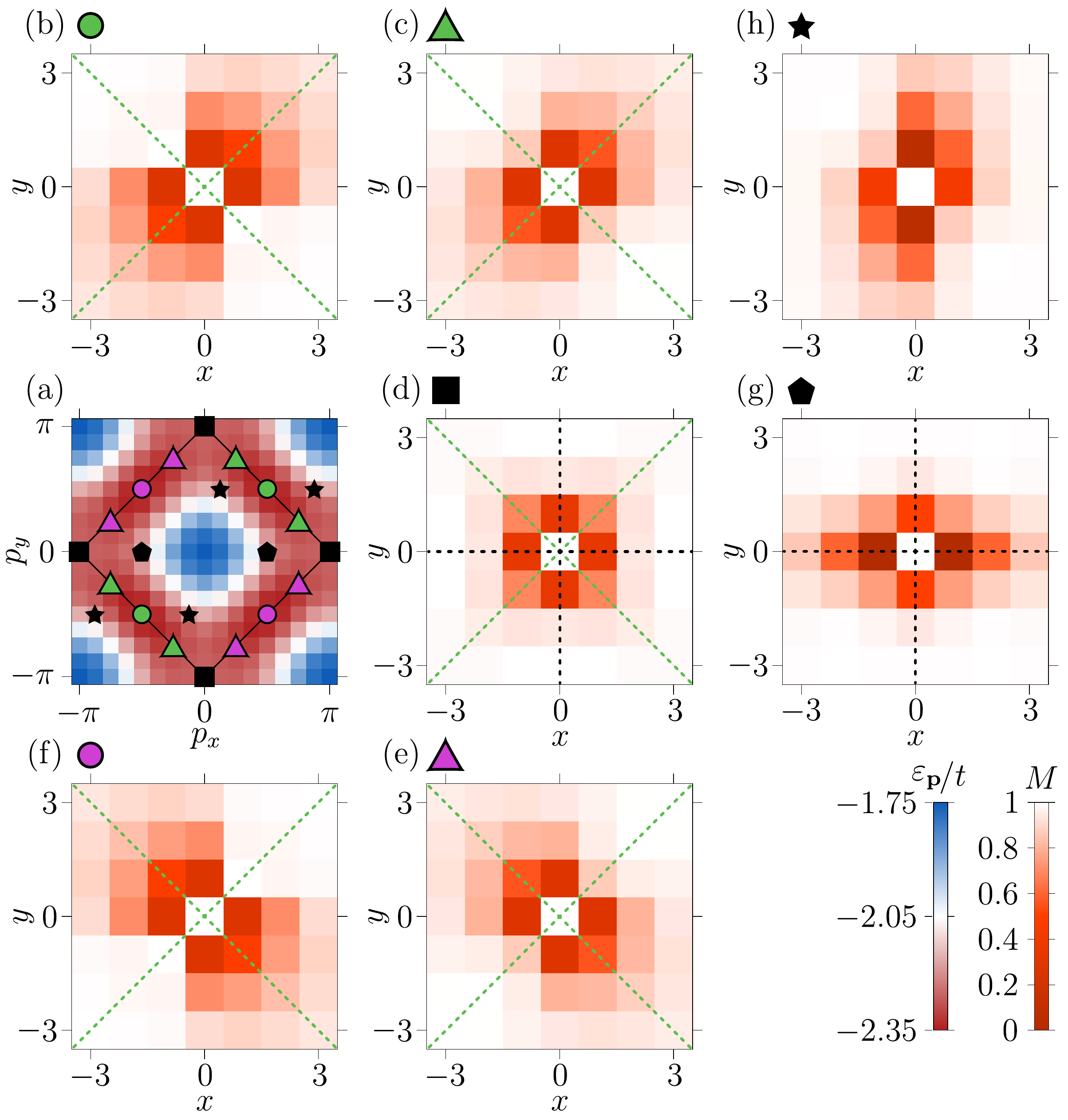}
\end{center}
\caption{Magnetization around the hole for $J / t = 0.3$ vs indicated crystal momenta [panel (a)] in the Heisenberg limit. For crystal momenta on the MBZ boundary [panels (b)--(f)], i.e., $|p_x| + |p_y| = \pi$, the system retains the reflection symmetries around the two diagonals (green-dashed lines), even though the total crystal momentum of the magnetic polaron might point off these axes. For crystal momenta marked with green (purple) symbols, the magnetization cloud is oriented along the $x = y$ ($x = -y$) diagonal. When $\bp=(\pm\pi,0), (0,\pm \pi)$ [black squares, panel (d)], the dressing cloud recovers the full $C_{4v}$ symmetry of the background antiferromagnetic order. This symmetry is again reduced when the crystal momentum does not lie on the MBZ boundary, as can be seen by comparing panels (d) [in the cases of $\bp = (\pm\pi,0)$] and (g) [$\bp = (\pm\pi/2,0)$], which have different symmetries even though the momenta are parallel. For a general momentum in panel (h), only the $C_2$ symmetry is retained and four crystal momentum states [black stars] show the same magnetization cloud pattern.}
\label{fig.magnetization_symmetries} 
\end{figure*} 

The symmetry group for the magnetic polaron must therefore be a descendant of $C_{4v}$. It turns out that the magnetic dressing cloud of the polaron has a remarkably high symmetry and that the full $C_{4v}$ is recovered for certain crystal momenta $\bp$ of the polaron. First, it follows from time-reversal symmetry that the dressing cloud is inversion symmetric, i.e. $M_\bp(\bd)=M_\bp(-\bd)$, for all crystal momenta, corresponding to the $C_2$ point symmetry group in two dimensions. The two-point hole-spin operator $\hat{M}^{(2)}(\bd) \propto \hat{h}^\dagger_{\mathbf 0}\hat{h}_{\mathbf 0}\hat{s}^\dagger_{\bd} \hat{s}_{\bd}$ [see Eq.\ \eqref{eq.M_2}] gives the spatial structure of the magnetization around the hole, and we can therefore 
argue for the spatial symmetries from here. Since it is hermitian, it fulfills the identity $\bra{\Psi_\bp}\hat{M}^{(2)}(\bd)\ket{\Psi_\bp} = \bra{\tilde{\Psi}_\bp} \hat{\mathcal{T}} \hat{M}^{(2)}(\bd) \hat{\mathcal{T}}^{-1} \ket{\tilde{\Psi}_\bp}$ \cite{Sakurai1994}. Here, $\hat{\mathcal{T}}$ is the anti-unitary time-reversal operator and $\ket{\tilde{\Psi}_\bp} = \hat{\mathcal{T}}\ket{\Psi_\bp}$ is the time-reversed polaron state. Now, the magnetization operator is invariant under $\hat{\mathcal{T}}$, i.e.\ $\hat{\mathcal{T}} \hat{M}^{(2)}(\bd) \hat{\mathcal{T}}^{-1} = \hat{M}^{(2)}(\bd)$, since reversal of time does not affect position operators. On the other hand, reversal of time flips the crystal momentum of the polaron so that $\hat{\mathcal{T}}\ket{\Psi_\bp} = \ket{\Psi_{-\bp}}$. Consequently, $M_{-\bp}(\bd) = M_\bp(\bd)$ and using the total inversion symmetry of the system $M_{-\bp}(\bd) = M_{\bp}(-\bd)$, we finally arrive at the $C_2$ inversion symmetry of the magnetization for any crystal momentum
\begin{align}
M_\bp(-\bd) = M_{\bp}(\bd).
\label{eq.general_C2_symmetry}
\end{align}
Note that this symmetry holds in any state given by a real linear combination of the crystal momentum eigenstates. This general $C_2$ point symmetry group of the magnetic dressing cloud may come as a surprise, since $\bp$ defines a specific direction which stands at odds with inversion symmetry along the direction of the crystal momentum. 

\begin{figure*}[t]
 \begin{center}
  \includegraphics[width=2\columnwidth]{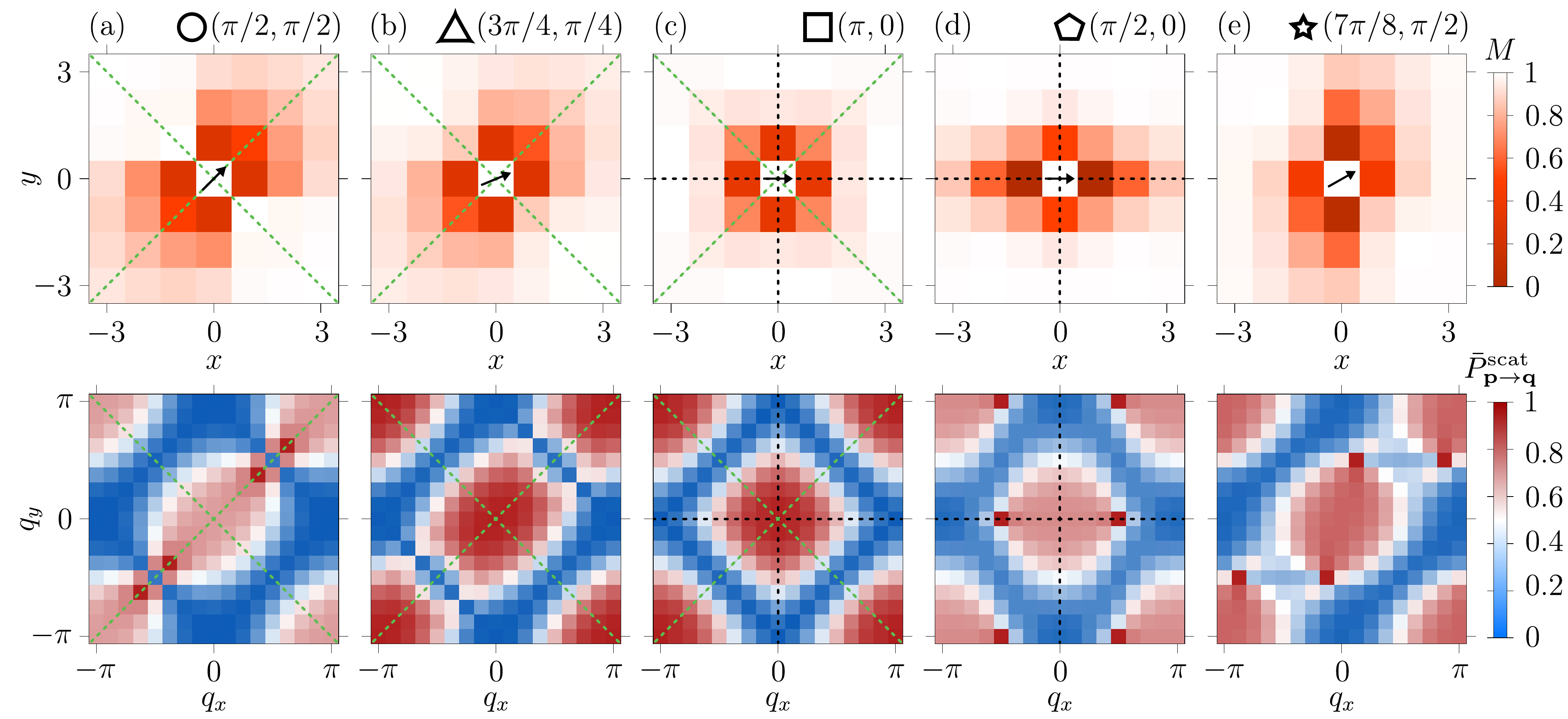}
 \end{center}
 \caption{Magnetization around the hole for $J / t = 0.3$ and different crystal momenta in the Heisenberg limit (top) compared to the scattering probability $\bar{P}^{\rm scat}_{\bp \to \bq} = P^{\rm scat}_{\bp \to +\bq} + P^{\rm scat}_{\bp \to -\bq}$ defined in Eq. \eqref{eq.scattering_probability} and normalized to its maximal value (bottom). While the crystal momentum of the magnetic polaron may be misaligned with the magnetization cloud, we find that the scattering probability, characterizing in which direction the hole preferably moves under the emission of spin waves, generally reflects the orientation of the magnetic dressing cloud. For the cases illustrated in panels (d) and (e), the polaron shows a strong tendency to scatter to very few momentum states, making the scattering profile remarkably sharp. }
 \label{fig.magnetization_vs_scattering_probability} 
\end{figure*} 

Higher spatial symmetries emerge for special crystal momenta. When $\bp$ is parallel to one of the lattice axes or one of the diagonals, the magnetization is symmetric under reflection operations parallel or perpendicular to the momentum. Combined with the general $C_2$ symmetry this forms the $C_{2v}$ symmetry group. Remarkably, the reflection symmetries along the lattice diagonals are retained for \emph{all} crystal momenta along the edge of the magnetic Brillouin zone (MBZ) given by $|p_x| + |p_y| = \pi$, as indicated by black lines in Fig. \ref{fig.magnetization_symmetries}(a). This can be understood from the symmetry of the magnetization under the translation $\bp \to \bp \pm \bQ$, where $\bQ = (\pi, \pi)$ is the wave vector of the antiferromagnetic spin-density wave. For crystal momenta along the line $(p_x, p_y)=(p_x, \pi - p_x)$, it follows that states with momenta $(p_x, \pi - p_x) - \bQ = (p_x - \pi , -p_x) = (-p_y, -p_x)$ must show the same magnetization pattern. This gives the reflection symmetry around the $y = -x$ diagonal. Using the $C_2$ symmetry described above then gives that $(p_y, p_x)$ and $(p_x,p_y)$ show the same magnetization pattern, which corresponds to a reflection symmetry around the $y = x$ diagonal. As a result, the dressing cloud of the magnetic polaron with momenta along the edge of the MBZ is characterized by the high point symmetry group $C_{2v}$. It even follows that when the momentum is at the corners of the MBZ, i.e.\ $\bp=(\pm\pi,0)$ or $\bp=(0,\pm\pi)$, the dressing cloud of the polaron has the full symmetry $C_{4v}$ of the AF state without the polaron. We note that this unusual spatial symmetry of the moving polaron is a fundamental property of the system, and holds generally irrespective of our SCBA treatment. Remarkably, the highest spatial symmetry of the magnetic polaron does, hereby, not emerge in its ground state at $\bp = (\pm \pi / 2, \pm \pi / 2)$.

The described symmetries are summarized in Fig. \ref{fig.magnetization_symmetries} showing the magnetic dressing cloud for different crystal momenta. Moving along the MBZ [Figs. \ref{fig.magnetization_symmetries}(b)--\ref{fig.magnetization_symmetries}(f)], the magnetization cloud undergoes a discrete rotation where it is always oriented along one of the diagonals, which leads to a misalignment of the dressing cloud and the crystal momentum. The dressing cloud recovers the full $C_{4v}$ symmetry of the background antiferromagnetic order when $\bp=(\pm\pi,0), (0,\pm \pi)$ [Fig. \ref{fig.magnetization_symmetries}(d)], whereas the symmetry descends to $C_{2v}$ when the momentum reduces to $\bp=(\pm\pi/2,0)$ [Fig. \ref{fig.magnetization_symmetries}(g)]. Finally, Fig. \ref{fig.magnetization_symmetries}(h) shows how the magnetic dressing cloud exhibits the minimal $C_2$ symmetry for a general momentum, while still being misaligned with the 
crystal momentum of the hole.

We can develop a microscopic picture of the discussed symmetries by considering the probability to find a hole at momentum $\bq$ in a polaron state with momentum $\bp$. This probability is given by
\begin{align}
P^{\rm scat}_{\bp \to \bq} = \bra{\Psi_\bp}\hat{h}^\dagger_{\bq}\hat{h}_{\bq}\ket{\Psi_\bp} - Z_\bp \cdot \delta_{\bp, \bq},
\label{eq.scattering_probability}
\end{align}
and describes the scattering between the hole and excitations of the AF in the magnetic polaron state. Since the bare hole state $\hat{h}_\bp^\dagger \ket{\AF}$ carries no spin wave excitations, and therefore does not contain any information about the magnetization cloud, we subtract the probability, $Z_\bp$, of remaining in that state. By itself, this scattering probability is not symmetric under inversion of the scattered momentum, $\bq\to - \bq$, but we can consider its symmetrized form $\bar{P}^{\rm scat}_{\bp \to \bq} = P^{\rm scat}_{\bp \to +\bq} + P^{\rm scat}_{\bp \to -\bq}$ which is shown in Fig. \ref{fig.magnetization_vs_scattering_probability} for selected crystal momenta. The depicted momentum distributions indicate the preferred directionality of the motion of the hole for a given momentum $\bp$ of the magnetic polaron, and its maxima indeed reflect the orientation of the magnetization cloud. In particular, for the ground state momentum, $\bp = (\pi / 2, \pi / 2)$, we see in Fig. \ref{fig.magnetization_vs_scattering_probability}(a) that the hole motion predominantly remains along the $x = y$ diagonal, thereby reducing the magnetization the most in this direction. The comparison between $M_{\bp}({\bf d})$ and $\bar{P}^{\rm scat}_{\bp \to \bq}$ for $\bp$ clearly shows that it is not the polaron crystal momentum but the preferred direction of the hole momentum that determines the symmetry and orientation of of the magnetic dressing cloud. 
Like the magnetization, $\bar{P}^{\rm scat}_{\bp \to \bq}$ is obtained from a self-consistency equation, which we derive in Appendix \ref{app.scattering_probability}. 

\subsection{Anisotropic spin interactions and the Ising limit} \label{subsec.heisenberg_ising_transition}
The developed framework also permits to study the transition from the Heisenberg to the Ising limit, by tuning the parameter $\alpha$ from $1$ to $0$ in Eq. \eqref{eq.H_J}. While choosing $\alpha\neq 1$ looses the correspondence with the original Fermi-Hubbard Hamiltonian, approaching the Ising limit makes the $t$-$J$ model better accessible to approximate treatments, such as a description in terms of defect-strings described in \cite{Bulaevskii1968,Brinkman1970,Trugman1988,Manousakis2007,Grusdt2018,Grusdt2018_2,Grusdt2019,Bohrdt2019}.

\begin{figure}[tb]
 \begin{center}
  \includegraphics[width=\columnwidth]{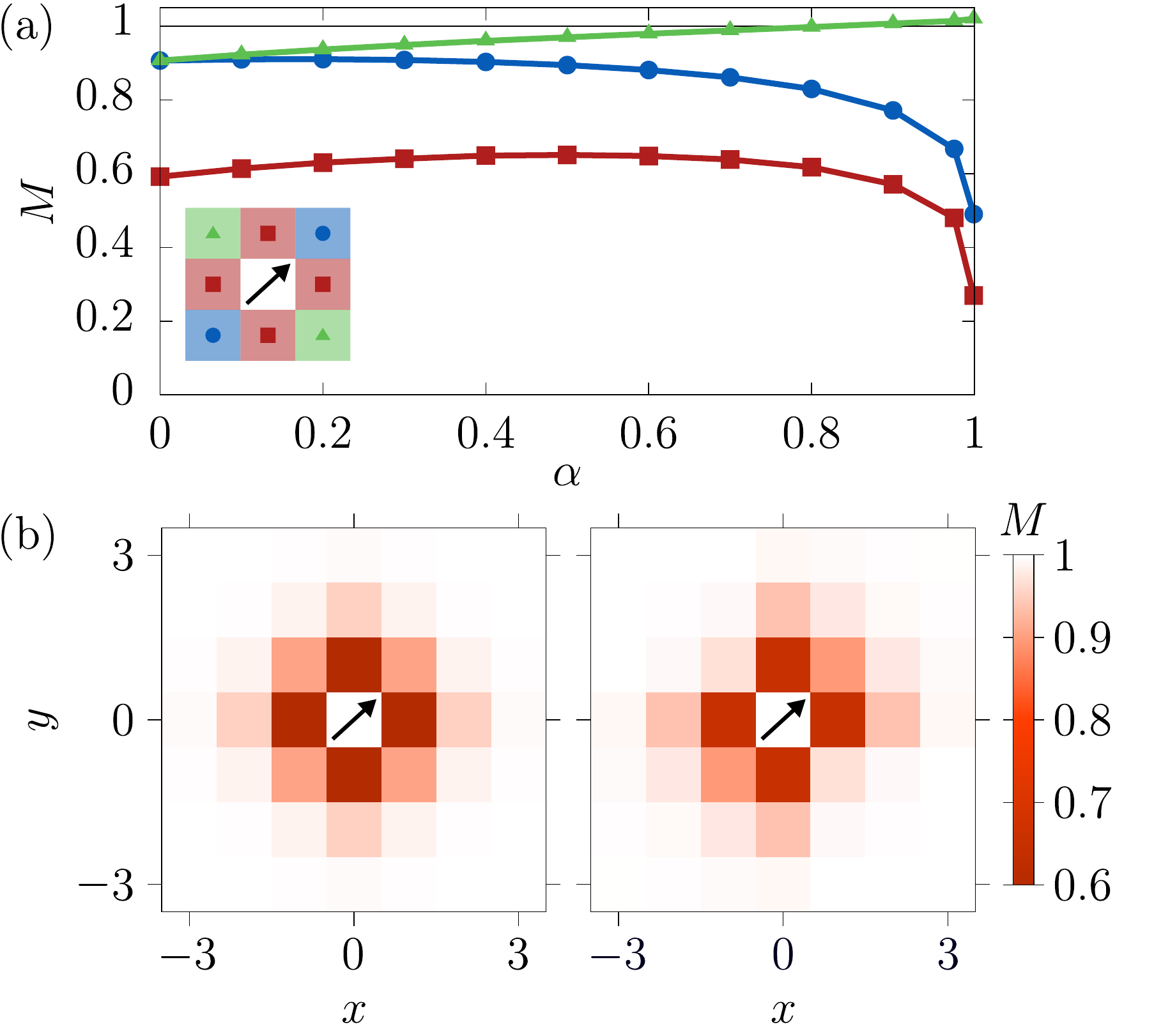}
 \end{center}
 \caption{Magnetization in the neighborhood of a hole for the polaron ground state with $\bp = (\pi / 2, \pi / 2)$ as a function of $\alpha$ determining the anisotropy of the spin-spin interactions for $J / t = 0.3$ in a 16 by 16 square. (a) Near the Heisenberg limit, $\alpha \lesssim 1$, the magnetization of the nearest neighbors (red squares) and of the next-nearest neighbors in direction of the hole crystal momentum (blue dots) rapidly changes with $\alpha$, whereas it only changes slightly for $0\leq\alpha \lesssim 0.8$. (b) Magnetization heat maps for the Ising case ($\alpha = 0$, left) and for $\alpha = 0.5$ (right). Note the difference in scale from Fig. \ref{fig.magnetization_vs_scattering_probability}(a), which illustrates the shrinking dressing cloud as the Ising limit is approached. }
 \label{fig.magnetization_alpha_vary} 
\end{figure}

In Fig. \ref{fig.magnetization_alpha_vary}(a), we plot the local magnetization around the hole as a function of $\alpha$ for $J / t = 0.3$. As one approaches the Ising limit, the magnetization cloud deforms and becomes more symmetric. This is associated with the appearance of a gap in the spin wave dispersion $\omega_\bk = zJ \sqrt{1 - \alpha^2 \gamma_\bk^2} / 2$, making it harder for the hole to emit spin waves when $\alpha < 1$. The direction perpendicular to $\bp$ stands out since the magnetization decreases as the Ising limit is approached, and the coherent increase of the magnetization in a Heisenberg magnet is lost. The Ising limit, $\alpha = 0$, restores the $C_{4v}$ symmetry of the magnetization, whereby the magnetization of all next nearest neighbors becomes identical.

This isotropy, shown in Fig.\ \ref{fig.magnetization_alpha_vary}(b), makes it possible to derive an analytical expression for the magnetization in the Ising limit, as we show explicitly in Appendix \ref{app.ising_limit}. Due to corrections to the hole Green's function beyond the SCBA \cite{Chernyshev1999} and the so-called Trugman loops \cite{Trugman1988}, this result is not exact. In fact, these small corrections have been shown \cite{Trugman1988,Chernyshev1999} to slightly lift the massive degeneracy of the crystal momentum states, favoring the crystal momentum $\bp = (0,0)$ as the ground state. Therefore, the reemergence of the $C_{4v}$ symmetry in the Ising limit \cite{Grusdt2018_2} can be expected to be an exact result for the ground state of the $t$-$J_z$ model. 

\section{Strong-coupling effects} \label{sec.non_perturbative_effects}
Previous SCBA calculations of spin-hole correlations have required a truncation of the SCBA polaron wave function restricting the number number of spin wave excitations \cite{Ramsak1998, Bala2000} or considered essentially flat excitation spectra \cite{Ramsak1998, Bala2002} to simplify higher-order terms in the polaron wave function. 

\begin{figure}[tb]
	\begin{center}
		\includegraphics[width=0.88\columnwidth]{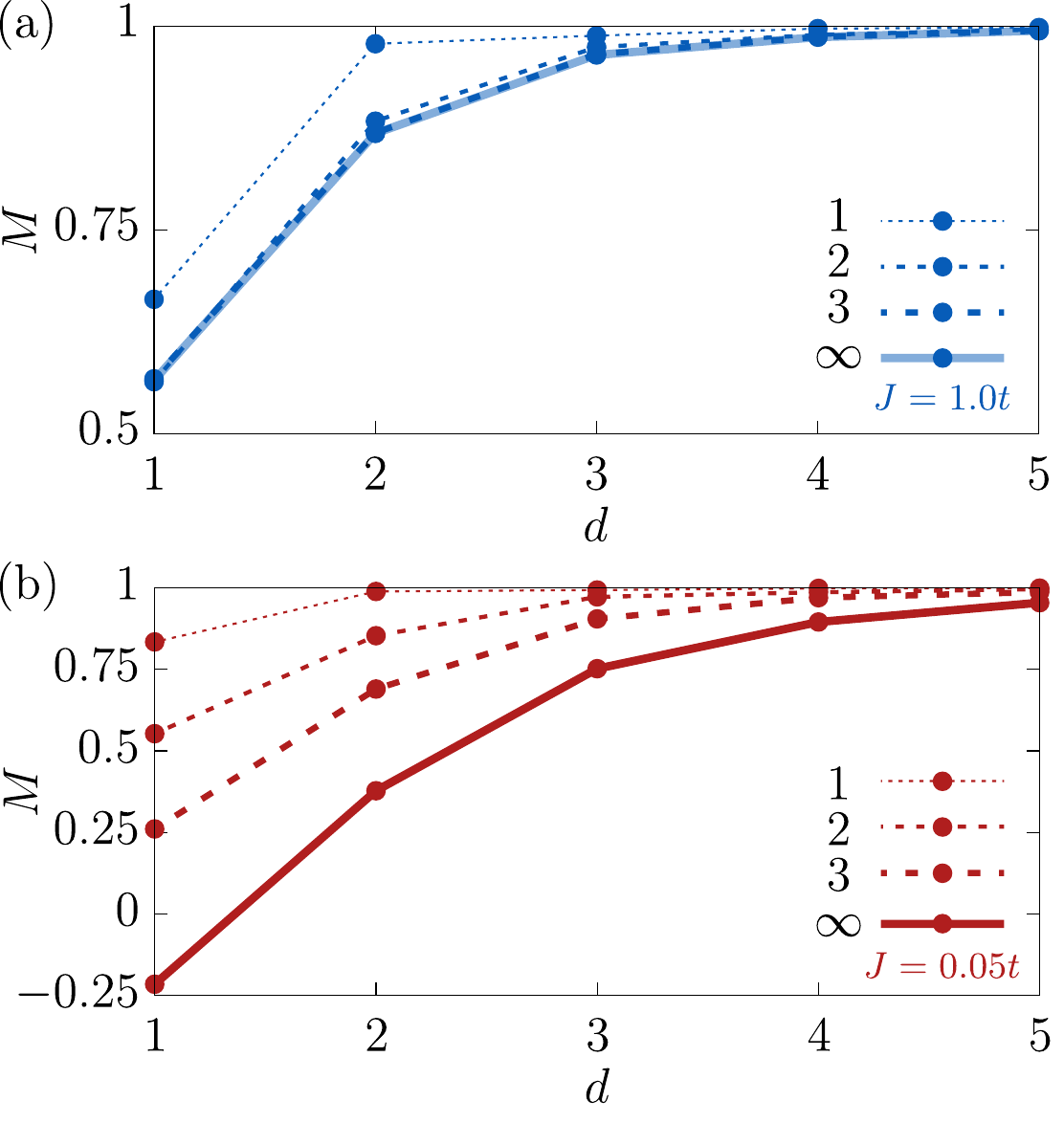}
	\end{center}
	\caption{Comparison of the full magnetization obtained from our non-perturbative calculation (solid lines) with that obtained by truncating the summation of the lowest terms (dashed lines) for $J = t$ (a) and $J = 0.05t$ (b) as a function of distance $d$ from the hole along the $y$-direction in a 16 by 16 square lattice. The dashed lines correspond to including 1, 2, and 3 spin waves in the magnetic polaron wave function, Eq. \eqref{eq.Reiters_wave_function}. For $J = t$, keeping only 2 spin waves gives an accurate description, while the 3 spin wave result is indistinguishable from the full calculation. For $J = 0.05t$, however, the full calculation yields quantitatively and qualitatively different results, including a sign flip of the magnetization at the nearest neighbor, $d = 1$. }
	\label{fig.magnetization_convergence} 
\end{figure}

Since the hole-spin interaction vertex scales as $g(\bq,\bp)\sim t$, and the spin wave energy scales as $\sim J$, such a truncation of the wave function can be understood as a perturbative series in $t/J$. This naturally limits the accuracy of such calculations to the weak to intermediate coupling regime $J/t\gtrsim 1$. This is illustrated in Fig. \ref{fig.magnetization_convergence}, where we compare the results from this truncation approach to our non-perturbative theory. For the interaction strength $J / t = 1$, the results converge nicely, and the magnetization calculated by including up to three spin waves in the wave function is essentially identical to the non-perturbative result. However, we see that the effects of the hole on the surrounding magnetization is significantly underestimated by the truncated wave functions for $J/t=0.05$. In particular, it completely misses the sign flip in the magnetization at the nearest neighbor sites. 

One should note that the $t$-$J$ model no longer describes the Fermi-Hubbard model when $J\sim t$. The found discrepancies, therefore, render any such truncation procedures practically inapplicable when comparing to the Fermi-Hubbard model. In contrast, the non-perturbative framework developed here allows for the inclusion of all terms in Eq. \eqref{eq.Reiters_wave_function}, whereby one can describe the spatial correlation of holes and spins deep in the strongly correlated regime of $J / t \ll 1$, in which the SCBA is expected to yield an efficient and accurate description of the $t$-$J$ Hamiltonian as well as the Fermi-Hubbard model around half filling. 

\section{Derivation of self-consistency equations} \label{sec.derivation}
We now describe the theoretical framework to include all terms in the SCBA wave function, Eq. \eqref{eq.Reiters_wave_function}, as used in the previous section. In particular, we derive self-consistency equations for the $B$ and $C$ functions in Eqs. \eqref{eq.B_position_space} and \eqref{eq.C_position_space}, which makes it possible to sum the infinite series in terms of spin wave excitation numbers. 

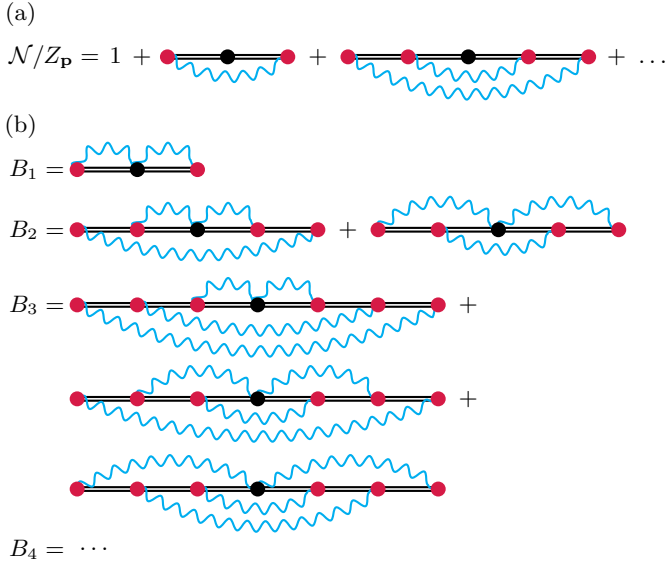
\begin{figure}[t!] 
\center
\begin{tikzpicture}[node distance=0.5cm and 1.5cm]
\coordinate[label = {(a)}] (a1);
\coordinate[below = 0.3cm of a1, label = right:{$\!\!\!\!\!\mathcal{N} / Z_\bp = $}] (b1);

\coordinate[right = 0.75cm of b1] (a2);
\coordinate[right = 0.8cm of a2, label=center:{$1 \;\, + \;\;$}] (b2);

\coordinate[right = 0.4cm of b2] (a3);
\coordinate[right = 0.8cm of a3] (b3);
\coordinate[right = 0.8cm of b3] (c3);

\coordinate[right = 0.4cm of c3, label=center:{$+$}] (d3);

\draw[thick, double] (a3) -- (b3);
\draw[thick, double] (b3) -- (c3);

\draw[thick, spinwave_total,cyan] (a3) to[out=-30,in=-150] (c3);

\foreach \n in {b3}
\node at (\n)[circle, fill, inner sep = 2pt, color = black]{};

\foreach \n in {a3,c3}
\node at (\n)[circle, fill, inner sep = 2pt, color = myred]{};

\coordinate[right = 0.4cm of d3] (a4);
\coordinate[right = 0.8cm of a4] (b4);
\coordinate[right = 0.8cm of b4] (c4);
\coordinate[right = 0.8cm of c4] (d4);
\coordinate[right = 0.8cm of d4] (e4);

\coordinate[right = 0.4cm of e4, label=center:{$\phantom{\;\dots}+\;\dots$}] (f4);

\draw[thick, double] (a4) -- (b4);
\draw[thick, double] (b4) -- (c4);
\draw[thick, double] (c4) -- (d4);
\draw[thick, double] (d4) -- (e4);

\draw[thick, spinwave_total,cyan] (a4) to[out=-30,in=-150] (e4);
\draw[thick, spinwave_total,cyan] (b4) to[out=-30,in=-150] (d4);

\foreach \n in {c4}
\node at (\n)[circle, fill, inner sep = 2pt, color = black]{};

\foreach \n in {a4,b4,d4,e4}
\node at (\n)[circle, fill, inner sep = 2pt, color = myred]{};

\coordinate[below = 1.5cm of a1, label = {(b)}] (a2);
\coordinate[below = 0.3cm of a2, label = right:{$\!\!\!\!\!B_1 = $}] (b2);

\coordinate[right = 0.75cm of b2] (c2);
\coordinate[right = 0.8cm of c2] (d2);
\coordinate[right = 0.8cm of d2] (e2);

\draw[thick, double] (c2) -- (d2);
\draw[thick, double] (d2) -- (e2);

\draw[thick,spinwave_total,cyan] (c2) to[out=90,in=90] (d2);
\draw[thick,spinwave_total,cyan] (d2) to[out=90,in=90] (e2);

\foreach \n in {d2}
\node at (\n)[circle, fill, inner sep = 2pt, color = black]{};

\foreach \n in {c2,e2}
\node at (\n)[circle, fill, inner sep = 2pt, color = myred]{};

\coordinate[below = 0.8cm of b2, label = right:{$\!\!\!\!\!B_2 = $}] (b3);
\coordinate[right = 0.75cm of b3] (c3);
\coordinate[right = 0.8cm of c3] (d3);
\coordinate[right = 0.8cm of d3] (e3);
\coordinate[right = 0.8cm of e3] (f3);
\coordinate[right = 0.8cm of f3] (g3);
\coordinate[right = 0.4cm of g3, label=center:{$+$}] (h3);

\draw[thick, double] (c3) -- (d3);
\draw[thick, double] (d3) -- (e3);
\draw[thick, double] (e3) -- (f3);
\draw[thick, double] (f3) -- (g3);

\draw[thick,spinwave_total,cyan] (d3) to[out=90,in=90] (e3);
\draw[thick,spinwave_total,cyan] (e3) to[out=90,in=90] (f3);

\draw[thick,spinwave_total,cyan] (c3) to[out=-20,in=-160] (g3);

\foreach \n in {e3}
\node at (\n)[circle, fill, inner sep = 2pt, color = black]{};

\foreach \n in {c3,d3,f3,g3}
\node at (\n)[circle, fill, inner sep = 2pt, color = myred]{};

\coordinate[right = 0.4cm of h3] (b4);
\coordinate[right = 0.8cm of b4] (c4);
\coordinate[right = 0.8cm of c4] (d4);
\coordinate[right = 0.8cm of d4] (e4);
\coordinate[right = 0.8cm of e4] (f4);

\draw[thick, double] (b4) -- (c4);
\draw[thick, double] (c4) -- (d4);
\draw[thick, double] (d4) -- (e4);
\draw[thick, double] (e4) -- (f4);

\draw[thick,spinwave_total,cyan] (b4) to[out=45,in=135] (d4);
\draw[thick,spinwave_total,cyan] (d4) to[out=45,in=135] (f4);

\draw[thick,spinwave_total,cyan] (c4) to[out=-30,in=-150] (e4);

\foreach \n in {d4}
\node at (\n)[circle, fill, inner sep = 2pt, color = black]{};

\foreach \n in {b4,c4,e4,f4}
\node at (\n)[circle, fill, inner sep = 2pt, color = myred]{};

\coordinate[below = 1cm of b3, label = right:{$\!\!\!\!\!B_3 = $}] (b5);
\coordinate[right = 0.75cm of b5] (c5);
\coordinate[right = 0.8cm of c5] (d5);
\coordinate[right = 0.8cm of d5] (e5);
\coordinate[right = 0.8cm of e5] (f5);
\coordinate[right = 0.8cm of f5] (g5);
\coordinate[right = 0.8cm of g5] (h5);
\coordinate[right = 0.8cm of h5] (i5);
\coordinate[right = 0.4cm of i5, label=center:{$+$}] (j5);

\draw[thick, double] (c5) -- (d5);
\draw[thick, double] (d5) -- (e5);
\draw[thick, double] (e5) -- (f5);
\draw[thick, double] (f5) -- (g5);
\draw[thick, double] (g5) -- (h5);
\draw[thick, double] (h5) -- (i5);

\draw[thick,spinwave_total,cyan] (e5) to[out=90,in=90] (f5);
\draw[thick,spinwave_total,cyan] (f5) to[out=90,in=90] (g5);

\draw[thick,spinwave_total,cyan] (c5) to[out=-25,in=-155] (i5);
\draw[thick,spinwave_total,cyan] (d5) to[out=-20,in=-160] (h5);

\foreach \n in {f5}
\node at (\n)[circle, fill, inner sep = 2pt, color = black]{};

\foreach \n in {c5,d5,e5,g5,h5,i5}
\node at (\n)[circle, fill, inner sep = 2pt, color = myred]{};

\coordinate[below = 1.25cm of c5] (c6);
\coordinate[right = 0.8cm of c6] (d6);
\coordinate[right = 0.8cm of d6] (e6);
\coordinate[right = 0.8cm of e6] (f6);
\coordinate[right = 0.8cm of f6] (g6);
\coordinate[right = 0.8cm of g6] (h6);
\coordinate[right = 0.8cm of h6] (i6);
\coordinate[right = 0.4cm of i6, label=center:{$+$}] (j6);

\draw[thick, double] (c6) -- (d6);
\draw[thick, double] (d6) -- (e6);
\draw[thick, double] (e6) -- (f6);
\draw[thick, double] (f6) -- (g6);
\draw[thick, double] (g6) -- (h6);
\draw[thick, double] (h6) -- (i6);

\draw[thick,spinwave_total,cyan] (d6) to[out=45,in=135] (f6);
\draw[thick,spinwave_total,cyan] (f6) to[out=45,in=135] (h6);

\draw[thick,spinwave_total,cyan] (c6) to[out=-20,in=-160] (i6);
\draw[thick,spinwave_total,cyan] (e6) to[out=-30,in=-150] (g6);

\foreach \n in {f6}
\node at (\n)[circle, fill, inner sep = 2pt, color = black]{};

\foreach \n in {c6,d6,e6,g6,h6,i6}
\node at (\n)[circle, fill, inner sep = 2pt, color = myred]{};

\coordinate[below = 1.2cm of c6] (c7);
\coordinate[right = 0.8cm of c7] (d7);
\coordinate[right = 0.8cm of d7] (e7);
\coordinate[right = 0.8cm of e7] (f7);
\coordinate[right = 0.8cm of f7] (g7);
\coordinate[right = 0.8cm of g7] (h7);
\coordinate[right = 0.8cm of h7] (i7);

\draw[thick, double] (c7) -- (d7);
\draw[thick, double] (d7) -- (e7);
\draw[thick, double] (e7) -- (f7);
\draw[thick, double] (f7) -- (g7);
\draw[thick, double] (g7) -- (h7);
\draw[thick, double] (h7) -- (i7);

\draw[thick,spinwave_total,cyan] (c7) to[out=30,in=150] (f7);
\draw[thick,spinwave_total,cyan] (f7) to[out=30,in=150] (i7);

\draw[thick,spinwave_total,cyan] (d7) to[out=-30,in=-150] (h7);
\draw[thick,spinwave_total,cyan] (e7) to[out=-30,in=-150] (g7);

\foreach \n in {f7}
\node at (\n)[circle, fill, inner sep = 2pt, color = black]{};

\foreach \n in {c7,d7,e7,g7,h7,i7}
\node at (\n)[circle, fill, inner sep = 2pt, color = myred]{};

\coordinate[below = 3.25cm of b5, label = right:{$\!\!\!\!\!B_4 = $}] (b8);
\coordinate[right = 1cm of b8, label = center:{$\dots$}] (c8);

\end{tikzpicture}
\caption{Diagrammatic representation for the norm series (a) and the $B$ series (b). The $B$ series comes from inserting the basic diagram, $B_1$, in the norm series (a). The shown diagrams are the ones that give a nonzero contribution to the $B$ function. These come in an increasing order denoted $B_n$, where $n$ denotes the number of interaction vertices (red dots) to each side. At a given order $n$, there are $n$ nonzero diagrams.}
\label{fig.B_series_terms}
\end{figure}

Figure \ref{fig.B_series_terms} shows the first few diagrams in the $B$ series. We construct these diagrams in the following way. First, we take the polaron wave function $\ket{\Psi_\bp}$ and its adjoint $\bra{\Psi_\bp}$, corresponding to Fig. \ref{fig.quasiparticle_properties}(b) and its mirror image respectively. Second, we place the operator $\hat{B} = N\sum_\bk \hat{h}^\dagger_{\bk + \bq_1}\hat{h}_{\bk + \bq_2} \hat{b}^\dagger_{-\bq_1} \hat{b}_{-\bq_2}$ between $\bra{\Psi_\bp}$ and $\ket{\Psi_\bp}$ to compute the expectation value $B(\bq_1, \bq_2; \bp, \varepsilon_\bp)$ in Eq. \eqref{eq.B_position_space}. A non-zero contribution to this expectation value involves the annihilation of a hole and a spin wave for both $\bra{\Psi_\bp}$ and $\ket{\Psi_\bp}$, which yields the structure $B_1$ in Fig. \ref{fig.B_series_terms}(b). Finally, all spin waves from $\bra{\Psi_\bp}$ and $\ket{\Psi_\bp}$ that are not annihilated by $\hat{B}$ must be joined together. Visually, this means that the series can be constructed from the norm series shown in Fig. \ref{fig.B_series_terms}(a). Specifically, we detach all double lines from black and red dots, insert the likewise detached $B_1$ diagram and reassemble the double lines with the dots. As a result, an infinite series of terms $B_1, B_2, \dots$ emerges \cite{Ramsak1993, Ramsak1998}, as shown in Fig. \ref{fig.B_series_terms}(b). The $n$-th term, $B_n = \sum_{i=1}^{n} B_n^{(i)}$, contains $n$ nonzero diagrams coming from the $n$ spin wave term in the polaron wave function \eqref{eq.Reiters_wave_function}. In all diagrams, we have suppressed the two overall single lines corresponding to the residue, $Z_\bp$, as depicted in Fig. \ref{fig.quasiparticle_properties}(e). 
All diagrams, where the spin wave lines of the wave function cross are not allowed within the SCBA, and have to be omitted for consistency. This is at the heart of the SCBA, in which only rainbow diagrams are included. Also, all diagrams that are left-right asymmetric vanish, as we show explicitly in Appendix \ref{app.vanishing_diagrams}. 

\begin{figure}[t!]
\center
\begin{tikzpicture}[node distance=0.5cm and 1.5cm]
\coordinate[label = {(a)}] (a1);

\coordinate[below = 0.3cm of a1, label = center:{$\;B_0$}] (b1);
\coordinate[right = 0.4cm of b1, label = center:{$=$}] (b1_2);
\coordinate[right = 0.4cm of b1_2] (c1);
\coordinate[right = 0.8cm of c1] (d1);
\coordinate[right = 0.8cm of d1] (e1);
\coordinate[right = 0.4cm of e1, label = center:{$+$}] (f1);

\draw[thick, double] (c1) -- (d1);
\draw[thick, double] (d1) -- (e1);

\draw[thick,spinwave_total,cyan] (c1) to[out=90,in=90] (d1);
\draw[thick,spinwave_total,cyan] (d1) to[out=90,in=90] (e1);

\foreach \n in {d1}
\node at (\n)[circle, fill, inner sep = 2pt, color = black]{};

\foreach \n in {c1,e1}
\node at (\n)[circle, fill, inner sep = 2pt, color = myred]{};

\coordinate[right = 0.4cm of f1] (a2);
\coordinate[right = 0.8cm of a2] (b2);
\coordinate[right = 0.8cm of b2] (c2);
\coordinate[right = 0.8cm of c2] (d2);
\coordinate[right = 0.8cm of d2] (e2);

\draw[thick, double] (a2) -- (b2);
\draw[thick, double] (b2) -- (c2);
\draw[thick, double] (c2) -- (d2);
\draw[thick, double] (d2) -- (e2);

\draw[thick,spinwave_total,cyan] (a2) to[out=45,in=135] (c2);
\draw[thick,spinwave_total,cyan] (c2) to[out=45,in=135] (e2);
\draw[thick,spinwave_total,cyan,double] (b2) to[out=-30,in=-150] (d2);

\foreach \n in {a2,b2,d2,e2}
\node at (\n)[circle, fill, inner sep = 2pt, color = myred]{};

\foreach \n in {c2}
\node at (\n)[circle, fill, inner sep = 2pt, color = black]{};

\coordinate[below = 1.25cm of a1, label = {(b)}] (a3_1);

\coordinate[below = 0.3cm of a3_1] (a3);
\coordinate[right = 0.7cm of a3] (b3);
\coordinate[right = 0.2cm of b3] (c3);
\coordinate[right = 0.7cm of c3] (d3);
\coordinate[right = 0.4cm of d3, label=center:{$=$}] (e3);

\coordinate[right = 0.1cm of b3] (c3_1);
\coordinate[above = 0.3cm of c3_1, label=center:{$F_B$}] (c3_2);

\draw[thick,double] (a3) -- (b3);
\draw[thick, double] (c3) -- (d3);

\draw[thick,spinwave_total,cyan,double] (a3) to[out=-30,in=-150] (d3);

\foreach \n in {a3,d3}
\node at (\n)[circle, fill, inner sep = 2pt, color = myred]{};

\coordinate[right = 0.4cm of e3] (a4);

\coordinate[right = 0.7cm of a4] (b4);
\coordinate[right = 0.2cm of b4] (c4);
\coordinate[right = 0.7cm of c4] (d4);
\coordinate[right = 0.4cm of d4, label=center:{$+$}] (e4);

\draw[thick,double] (a4) -- (b4);
\draw[thick, double] (c4) -- (d4);

\draw[thick,spinwave_total,cyan] (a4) to[out=-30,in=-150] (d4);

\foreach \n in {a4,d4}
\node at (\n)[circle, fill, inner sep = 2pt, color = myred]{}; 

\coordinate[right = 0.4cm of e4] (a5);

\coordinate[right = 0.8cm of a5] (b5);
\coordinate[right = 0.7cm of b5] (c5);
\coordinate[right = 0.2cm of c5] (d5);
\coordinate[right = 0.7cm of d5] (e5);
\coordinate[right = 0.8cm of e5] (f5);

\coordinate[right = 0.1cm of c5] (c5_1);
\coordinate[above = 0.3cm of c5_1, label=center:{$F_B$}] (c5_2);

\draw[thick,double] (a5) -- (b5);
\draw[thick,double] (b5) -- (c5);

\draw[thick,double] (d5) -- (e5);
\draw[thick,double] (e5) -- (f5);

\draw[thick,spinwave_total,cyan] (a5) to[out=-30,in=-150] (f5);
\draw[thick,spinwave_total,cyan,double] (b5) to[out=-30,in=-150] (e5);

\foreach \n in {a5,b5,e5,f5}
\node at (\n)[circle, fill, inner sep = 2pt, color = myred]{};

\coordinate[below = 1.5cm of a3_1, label = {(c)}] (a6);

\coordinate[below = 0.3cm of a6, label = center:{$\;B\phantom{_0}$}] (b6);
\coordinate[right = 0.4cm of b6, label = center:{$=$}] (c6);
\coordinate[right = 0.65cm of c6, label = center: {$B_0 \; + \; $}] (d6);

\coordinate[right = 0.5cm of d6] (a7);
\coordinate[right = 0.7cm of a7] (b7);
\coordinate[right = 0.35cm of b7, label=center:{$B_0$}] (B0_center_2);
\coordinate[right = 0.35cm of B0_center_2] (c7);
\coordinate[right = 0.7cm of c7] (d7);
\coordinate[right = 0.4cm of d7, label = center:{$\phantom{\dots}+\dots$}] (e7);

\draw[thick,double] (a7) -- (b7);
\draw (B0_center_2) circle (0.35cm);
\draw[thick,double] (c7) -- (d7);

\draw[spinwave_total,thick,cyan] (a7) to[out=-45,in=-135] (d7);

\foreach \n in {a7,d7}
\node at (\n)[circle, fill, inner sep = 2pt, color = myred]{};

\coordinate[right = 0.75cm of e7, label = center:{$=$}] (a8);
\coordinate[right = 0.65cm of a8, label = center: {$B_0 \; + \; $}] (b8);

\coordinate[right = 0.5cm of b8] (a9);
\coordinate[right = 0.7cm of a9] (b9);
\coordinate[right = 0.35cm of b9, label=center:{$B$}] (B_center);
\coordinate[right = 0.35cm of B_center] (c9);
\coordinate[right = 0.7cm of c9] (d9);

\draw[thick,double] (a9) -- (b9);
\draw (B_center) circle (0.35cm);
\draw[thick,double] (c9) -- (d9);

\draw[spinwave_total,thick,cyan] (a9) to[out=-45,in=-135] (d9);

\foreach \n in {a9,d9}
\node at (\n)[circle, fill, inner sep = 2pt, color = myred]{};

\end{tikzpicture}
\caption{(a) Summation of all the last diagrams, $B_0 = \sum_{n} B_{n}^{(n)}$. (b) The $F_B$ function thus emerges. (c) Finally, the full $B$ function comes about by putting $0, 1, 2 \dots$ spin wave lines around $B_0$. This results in a self-consistent equation for the $B$ function. }
\label{fig.B0_and_B}
\end{figure}
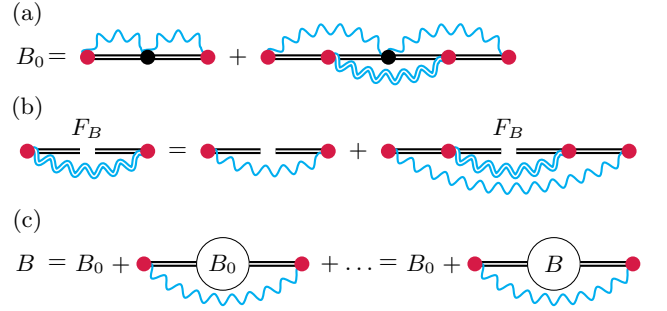

The central idea to obtain the self-consistency equations is the following. First, we take the last diagram from each order, $B_{n}^{(n)}$, and sum up only these, to obtain
\begin{equation}
B_0 = \sum_{n = 1}^{\infty} B_{n}^{(n)}.
\label{eq.B_0_definition}
\end{equation}
This leads to the diagrammatic structure shown in Figs. \ref{fig.B0_and_B}(a) and \ref{fig.B0_and_B}(b). Second, we notice that all other terms in Fig. \ref{fig.B_series_terms} are related to $B_0$ by 
putting a number of spin wave lines around $B_0$. The result is the full $B$ function shown in Fig. \ref{fig.B0_and_B}(c). Algebraically, the $B_0$ function is thus
\begin{align}
B_0(\bq_1, \bq_2; \bp, \omega) =&\; N g(\bp, \bq_1) g(\bp, \bq_2) \nn \\
& \cdot G(\bp \!+\! \bq_1, \omega \!-\! \omega_{\bq_1})G(\bp \!+\! \bq_2, \omega \!-\! \omega_{\bq_2})\nn \\
& \cdot \left[1 + F_B(\bq_1, \bq_2; \bp, \omega)\right].
\label{eq.B0_and_FB}
\end{align}
This is written in terms of the function $F_B$, which fulfills the self-consistency equation 
\begin{align}
\!\!\!\!\! F_B(\bq_1, \bq_2; \bp, \omega) =& \sum_\bk g(\bp \!+\! \bq_1, \bk)g(\bp \!+\! \bq_2, \bk) \nn \\
& \cdot G(\bp \!+\! \bq_1 \!+\! \bk, \omega \!-\! \omega_{\bq_1} \!-\! \omega_\bk) \nn \\
& \cdot G(\bp \!+\! \bq_2 \!+\! \bk, \omega \!-\! \omega_{\bq_2} \!-\! \omega_\bk) \nn \\
& \cdot \left[1 + F_B(\bq_1, \bq_2; \bp \!+\! \bk, \omega \!-\! \omega_\bk)\right],
\label{eq.FB}
\end{align}
that is depicted diagrammatically in Fig. \ref{fig.B0_and_B}(b). Both of these depend on the general energy $\omega \leq \varepsilon_\bp$. Finally, the self-consistency equation shown in Fig. \ref{fig.B0_and_B}(c) can be written as
\begin{align}
B(\bq_1, \bq_2; \bp, \omega) =&\; B_0(\bq_1, \bq_2; \bp, \omega) \nn \\
& + \sum_\bk g^2(\bp, \bk) G^2(\bp \!+\! \bk, \omega \!-\! \omega_\bk) \nn \\
& \cdot B(\bq_1, \bq_2; \bp \!+\! \bk, \omega \!-\! \omega_\bk).
\label{eq.self_consistency_equation_B}
\end{align}
Importantly, we can relate the structure of Eqs. \eqref{eq.FB} and \eqref{eq.self_consistency_equation_B} to that of the self-energy, in Eq. \eqref{eq.hole_self_energy_self_consistency_equation}. Specifically, taking the derivative of the self-energy equation yields
\begin{align}
-\partial_\omega \Sigma(\bp, \omega) =&\, \sum_\bk g^2(\bp, \bk) G^2(\bp \!+\! \bk, \omega \!-\!\omega_\bk) \nn \\
&\, \cdot \left[1 - \partial_\omega \Sigma(\bp \!+\! \bk, \omega \!-\! \omega_\bk)\right], 
\label{eq.derivative_self_energy}
\end{align}
which corresponds to the norm series in Fig. \ref{fig.B_series_terms}(a), apart from the first term. Comparing this to the self-consistency equation for the $B$ function, we see that $-\partial_\omega \Sigma(\bp, \omega)$ takes on the role of $B$, while $\sum_\bk g^2(\bp, \bk) G^2(\bp \!+\! \bk, \omega \!-\!\omega_\bk)$ corresponds to $B_0$. For each external momenta $\bq_1$ and $\bq_2$, these more advanced self-consistency equations, thus, show the exact same structure as that of the self-energy. In fact, computing $F_B({\mathbf 0}, {\mathbf 0}; \bp, \omega + \omega_{\mathbf 0})$ in Eq. \eqref{eq.FB} and comparing it to Eq. \eqref{eq.derivative_self_energy} shows that
\begin{align}
F_B({\mathbf 0}, {\mathbf 0}; \bp, \omega + \omega_{\mathbf 0}) = -\pa_\omega \Sigma(\bp, \omega), 
\label{eq.FB_and_Sigma_derivative}
\end{align}
which in turn is related to the quasiparticle residue in Eq. \eqref{eq.quasiparticle_residue}. This link provides a useful consistency check for the numerical calculations, and also offers an alternative way of computing the residue. Furthermore, it shows that the order of the $B$ and $C$ functions (including the overall factor of $Z_\bp$) is $Z_\bp \cdot \mathcal{O}[-\pa_\omega \Sigma(\bp, \omega)] = Z_\bp (1 / Z_\bp - 1) = 1 - Z_\bp$. This emphasizes that small quasiparticle residues, $Z_\bp \ll 1$, correspond to large changes in the local magnetization cloud, $B, C \sim 1$, as was discussed in more general terms in Sec. \ref{sec.magnetic_polaron_quasiparticle}. 

For the $C$ function, an equivalent derivation, see Appendix \ref{app.derivation_C_function}, leads to the feeding term
\begin{align}
C_0(\bq_1, \bq_2; \bp, \omega) = N \big[ & g(\bp, \bq_1) g(\bp \!+\! \bq_1, \bq_2) G(\bp \!+\! \bq_1, \omega \!-\! \omega_{\bq_1}) \nn \\
+ & g(\bp, \bq_2) g(\bp \!+\! \bq_2, \bq_1) G(\bp \!+\! \bq_2, \omega \!-\! \omega_{\bq_2}) \big] \nn \\
& \cdot G(\bp \!+\! \bq_1 \!+\! \bq_2, \omega \!-\! \omega_{\bq_1} \!-\! \omega_{\bq_2}), \nn \\
& \cdot \left[1 + F_C(\bq_1, \bq_2; \bp, \omega)\right]. 
\label{eq.C0_and_FC}
\end{align}
This is written in terms of the $F_C$ function 
\begin{align}
\!\!\!\!\! F_C(\bq_1, \bq_2; \bp, \omega) =& \sum_\bk g(\bp, \bk)g(\bp \!+\! \bq_1 \!+\! \bq_2, \bk) \nn \\
& \cdot G(\bp \!+\! \bk, \omega \!-\! \omega_\bk) \nn \\
& \cdot G(\bp \!+\! \bq_1 \!+\! \bq_2 \!+\! \bk, \omega \!-\! \omega_{\bq_1} \!-\! \omega_{\bq_2} \!-\! \omega_\bk) \nn \\
& \cdot \left[1 + F_C(\bq_1, \bq_2; \bp \!+\! \bk, \omega \!-\! \omega_\bk)\right],
\label{eq.FC}
\end{align}
which has the same role as $F_B$ in the $B$ series, and which it also closely resembles. The final self-consistency equation for the $C$ function is identical to Eq. \eqref{eq.self_consistency_equation_B} with $C$ and $C_0$ playing the role of $B$ and $B_0$ respectively.

The full computation of the magnetization in the neighborhood of a hole can then be performed as follows. For each pair of external spin wave momenta $\bq_1, \bq_2$, we first solve the two self-consistency equations for $F_B$ and $F_C$, Eqs. \eqref{eq.FB} and \eqref{eq.FC} respectively. From these, we calculate $B_0$ and $C_0$, Eqs. \eqref{eq.B0_and_FB} and \eqref{eq.C0_and_FC}. These are used in the self-consistency equation \eqref{eq.self_consistency_equation_B}. We then multiply $B$ and $C$ with the overall residue $Z_\bp$, and finally transform them to position space according to Eqs. \eqref{eq.B_position_space} and \eqref{eq.C_position_space}. 
The task of solving $N^2$ self-consistency equations may seem daunting at first sight. A major simplification, however, comes from a number of symmetry properties of the $B$ and $C$ functions in momentum space, as outlined in Appendix \ref{app.symmetries_of_B_and_C}. Therefore, only a small fraction of the $N^2$ equations has to be solved explicitly, whereby all results presented in this work could be obtained with modest computational resources.

\section{Experiments} \label{sec.experiments}
The spatial structure of magnetic polarons can play an important role for the transport properties of electrons in a solid. While experiments show evidence for their formation in the cuprates \cite{Wells1995,Ronning1998,Kim1998,Yoshida2003,Ronning2005,Rosch2005,highTc}, a detailed and direct probing of the underlying correlations on a microscopic level has not been possible in condensed matter measurements. This has changed with the development of quantum simulation platforms based on cold atoms in optical lattices \cite{2010Esslinger,Gross2017,Schafer2020}, and has in recent years made it possible to implement near-perfect realizations of the Fermi-Hubbard model \cite{Boll2016,Cheuk2016b,Mazurenko2017,Hilker2017,Brown2017,Chiu2018,Brown2019,Koepsell2019,Chiu2019,Koepsell2020,Brown2020a,Vijayan2020,Hartke2020,Brown2020b,Ji2021}. 

The single-site resolution achievable in current experiments permits to image any desired correlation function between particles, and in particular between a dopant and its surrounding effective spins. This opens up the possibility to probe the inner structure of the magnetic polaron and its motion, as explored in the present work. So far, the lowest achieved temperatures are around $k_{\rm B}T = 0.5 J$ \cite{Chiu2019}, at which there may be significant thermal corrections to the pure polaron states investigated here. Reaching lower temperatures will be an important step for direct comparisons and is widely expected to yield key insights into the microscopic physics of magnetic polarons and their role in high-$T_c$ superconductivity in strongly correlated materials. 

While such quantum simulators of the Fermi-Hubbard Hamiltonian naturally realize the isotropic $t$-$J$ model, the versatile toolbox to control and manipulate cold atoms also makes it possible to implement more general spin Hamiltonians, such as considered in Eq.\ \eqref{eq.H_J}. This includes polar molecules \cite{Gorshkov2011_2}, as well as Rydberg-dressed atoms in optical lattices \cite{Glaetzle2015,Bijnen2015,Zeiher2016,Zeiher2017,Borish2020,Sanchez2021}, which will make it possible to continuously tune between the $t$-$J$ and $t$-$J_z$ model, and to realize low temperatures compared to the much larger spin interactions achievable in these systems.

\section{Conclusions and outlook}
Inspired by recent experimental breakthroughs, we explored the properties of magnetic polarons that are formed by a hole and spin fluctuations in an antiferromagnetic square lattice, as described by the $t$-$J$ model. By combining the SCBA for the hole Green's function with the many-body wave function wave of the polaron, we developed a non-perturbative resummation scheme that now makes it possible to determine spin-hole correlations in the strongly interacting regime. This method thus enables broad explorations of the microscopic structure of magnetic polarons, which, so far, has not been possible within the SCBA. Given the proven accuracy of the SCBA for one-body observables such as the energy and residue of the polaron, this constitutes a significant step forward and will enable detailed analyses of ongoing experiments based on cold-atom quantum simulators. To illustrate the power of the approach, we have considered the magnetization in the vicinity of the hole, which turns out to deviate considerable from previous perturbative results under conditions where the $t$-$J$ model is valid. For a moving hole the magnetization cloud has an elongated shape, which features a surprising misalignment with the hole momentum that originates from the various symmetries of the antiferromagnetic state of the underlying spin lattice. 

The demonstrated possibility to explore correlations within the SCBA opens up new perspectives for studying strongly correlated quantum matter, including the physics of cuprates in the limit of small doping. It has been shown \cite{Kyung1996,Xiang1996,Manousakis2007} that these systems can be modeled quantitatively by including next-nearest neighbor hopping terms in the Hamiltonian, defining the so-called $t$-$t'$-$t''$-$J$ model. Within linear spin wave theory, this amounts to the addition of kinetic energy terms for the hole and can, thus, straightforwardly be including in our methodology.

The developed approach in general offers a promising starting point for extensions of the method along several directions and will facilitate detailed comparisons to recent cold-atom experiments on the Fermi-Hubbard model. While we have focused here on two-dimensional square lattices, the developed theoretical framework can be equally applied to any bipartite Bravais lattice in one, two or three spatial dimensions. It will moreover be interesting to assess corrections beyond linear spin wave theory \cite{Singh1995,Kim1999,Sandvik2001,Ronnow2001,Coldea2001,Christensen2004} and to explore higher-order correlation functions \cite{Grusdt2018_2,Grusdt2019,Blomquist2020}, which have been observed in recent cold-atom experiments \cite{Koepsell2019,Chiu2019,Koepsell2020}. The SCBA approach as used here to describe spin-hole correlations may also be employed to analyze correlations between two holes \cite{Dagotto1994,Chernyshev1994,Belinicher1997}, which will contribute to the understanding of pairing and a potential mechanism for high temperature superconductivity in the limit of small doping \cite{Schrieffer1988,Shraiman1989,Frenkel1990,Eder1992,Izyumov1997,Riera1998}. While the present pure-state treatment restricts our predictions to zero temperature, generalizing our framework to finite temperatures will make it possible to characterize the impact of temperatures that are currently achievable in optical-lattice experiments \cite{Koepsell2019,Chiu2019,Koepsell2020} and enable tests of the SCBA framework based on direct comparisons to measured correlation functions at finite temperature and strong interactions. Circumventing current temperature limitations, recent experiments \cite{Ji2021} have probed the transient dynamics following hole creation \cite{Bohrdt2020,Hubig2020}, which permits to trace the formation of magnetic polarons. We anticipate that the SCBA approach developed in this work can also provide an accurate framework to describe the non-equilibrium dynamics of polarons in strongly interacting quantum magnets.
 
\acknowledgments
We thank Annabelle Bohrdt and Fabian Grusdt for valuable feedback on our manuscript. KKN would like to thank Simon Panyella Pedersen for setting up the numerical calculation on the CSCAA Grendel cluster. This work has been supported by the Danish National Research Foundation through the Center of Excellence “CCQ” (Grant agreement no.: DNRF156).

\appendix

\section{Derivation of $C$ function} \label{app.derivation_C_function}
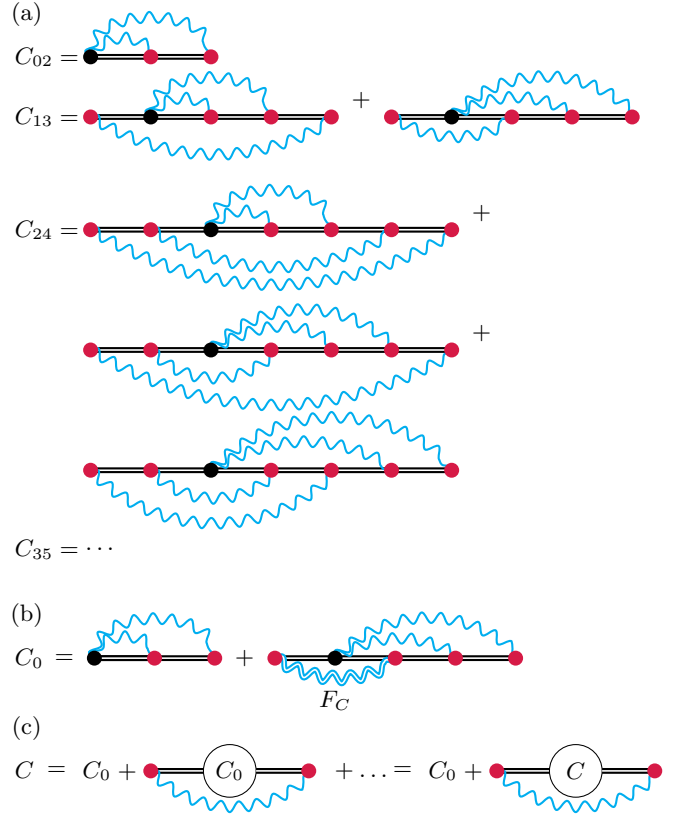
\begin{figure}[t!]
\center
\begin{tikzpicture}[node distance=0.5cm and 1.5cm]

\coordinate[label = {(a)}] (a2);
\coordinate[below = 0.3cm of a2, label = right:{$\!\!\!\!\!C_{02} = $}] (b2);

\coordinate[right = 0.85cm of b2] (c2);
\coordinate[right = 0.8cm of c2] (d2);
\coordinate[right = 0.8cm of d2] (e2);

\draw[thick, double] (c2) -- (d2);
\draw[thick, double] (d2) -- (e2);

\draw[thick,spinwave_total,cyan] (c2) to[out=90,in=90] (e2);
\draw[thick,spinwave_total,cyan] (c2) to[out=60,in=120] (d2);

\foreach \n in {c2}
\node at (\n)[circle, fill, inner sep = 2pt, color = black]{};

\foreach \n in {d2,e2}
\node at (\n)[circle, fill, inner sep = 2pt, color = myred]{};

\coordinate[below = 0.8cm of b2, label = right:{$\!\!\!\!\!C_{13} = $}] (b3);
\coordinate[right = 0.85cm of b3] (c3);
\coordinate[right = 0.8cm of c3] (d3);
\coordinate[right = 0.8cm of d3] (e3);
\coordinate[right = 0.8cm of e3] (f3);
\coordinate[right = 0.8cm of f3] (g3);
\coordinate[right = 0.4cm of g3, label={$+$}] (h3);

\draw[thick, double] (c3) -- (d3);
\draw[thick, double] (d3) -- (e3);
\draw[thick, double] (e3) -- (f3);
\draw[thick, double] (f3) -- (g3);

\draw[thick,spinwave_total,cyan] (d3) to[out=90,in=90] (f3);
\draw[thick,spinwave_total,cyan] (d3) to[out=60,in=120] (e3);

\draw[thick,spinwave_total,cyan] (c3) to[out=-30,in=-150] (g3);

\foreach \n in {d3}
\node at (\n)[circle, fill, inner sep = 2pt, color = black]{};

\foreach \n in {c3,e3,f3,g3}
\node at (\n)[circle, fill, inner sep = 2pt, color = myred]{};

\coordinate[right = 0.4cm of h3] (b4);
\coordinate[right = 0.8cm of b4] (c4);
\coordinate[right = 0.8cm of c4] (d4);
\coordinate[right = 0.8cm of d4] (e4);
\coordinate[right = 0.8cm of e4] (f4);

\draw[thick, double] (b4) -- (c4);
\draw[thick, double] (c4) -- (d4);
\draw[thick, double] (d4) -- (e4);
\draw[thick, double] (e4) -- (f4);

\draw[thick,spinwave_total,cyan] (c4) to[out=30,in=150] (e4);
\draw[thick,spinwave_total,cyan] (c4) to[out=45,in=135] (f4);

\draw[thick,spinwave_total,cyan] (b4) to[out=-30,in=-150] (d4);

\foreach \n in {c4}
\node at (\n)[circle, fill, inner sep = 2pt, color = black]{};

\foreach \n in {b4,d4,e4,f4}
\node at (\n)[circle, fill, inner sep = 2pt, color = myred]{};

\coordinate[below = 1.5cm of b3, label = right:{$\!\!\!\!\!C_{24} = $}] (b5);
\coordinate[right = 0.85cm of b5] (c5);
\coordinate[right = 0.8cm of c5] (d5);
\coordinate[right = 0.8cm of d5] (e5);
\coordinate[right = 0.8cm of e5] (f5);
\coordinate[right = 0.8cm of f5] (g5);
\coordinate[right = 0.8cm of g5] (h5);
\coordinate[right = 0.8cm of h5] (i5);
\coordinate[right = 0.4cm of i5, label={$+$}] (j5);

\draw[thick, double] (c5) -- (d5);
\draw[thick, double] (d5) -- (e5);
\draw[thick, double] (e5) -- (f5);
\draw[thick, double] (f5) -- (g5);
\draw[thick, double] (g5) -- (h5);
\draw[thick, double] (h5) -- (i5);

\draw[thick,spinwave_total,cyan] (e5) to[out=60,in=120] (f5);
\draw[thick,spinwave_total,cyan] (e5) to[out=90,in=90] (g5);

\draw[thick,spinwave_total,cyan] (c5) to[out=-30,in=-150] (i5);
\draw[thick,spinwave_total,cyan] (d5) to[out=-30,in=-150] (h5);

\foreach \n in {e5}
\node at (\n)[circle, fill, inner sep = 2pt, color = black]{};

\foreach \n in {c5,d5,f5,g5,h5,i5}
\node at (\n)[circle, fill, inner sep = 2pt, color = myred]{};

\coordinate[below = 1.6cm of c5] (c6);
\coordinate[right = 0.8cm of c6] (d6);
\coordinate[right = 0.8cm of d6] (e6);
\coordinate[right = 0.8cm of e6] (f6);
\coordinate[right = 0.8cm of f6] (g6);
\coordinate[right = 0.8cm of g6] (h6);
\coordinate[right = 0.8cm of h6] (i6);
\coordinate[right = 0.4cm of i6, label={$+$}] (j6);

\draw[thick, double] (c6) -- (d6);
\draw[thick, double] (d6) -- (e6);
\draw[thick, double] (e6) -- (f6);
\draw[thick, double] (f6) -- (g6);
\draw[thick, double] (g6) -- (h6);
\draw[thick, double] (h6) -- (i6);

\draw[thick,spinwave_total,cyan] (e6) to[out=30,in=135] (g6);
\draw[thick,spinwave_total,cyan] (e6) to[out=45,in=135] (h6);

\draw[thick,spinwave_total,cyan] (c6) to[out=-30,in=-150] (i6);
\draw[thick,spinwave_total,cyan] (d6) to[out=-45,in=-135] (f6);

\foreach \n in {e6}
\node at (\n)[circle, fill, inner sep = 2pt, color = black]{};

\foreach \n in {c6,d6,f6,g6,h6,i6}
\node at (\n)[circle, fill, inner sep = 2pt, color = myred]{};

\coordinate[below = 1.6cm of c6] (c7);
\coordinate[right = 0.8cm of c7] (d7);
\coordinate[right = 0.8cm of d7] (e7);
\coordinate[right = 0.8cm of e7] (f7);
\coordinate[right = 0.8cm of f7] (g7);
\coordinate[right = 0.8cm of g7] (h7);
\coordinate[right = 0.8cm of h7] (i7);

\draw[thick, double] (c7) -- (d7);
\draw[thick, double] (d7) -- (e7);
\draw[thick, double] (e7) -- (f7);
\draw[thick, double] (f7) -- (g7);
\draw[thick, double] (g7) -- (h7);
\draw[thick, double] (h7) -- (i7);

\draw[thick,spinwave_total,cyan] (e7) to[out=30,in=150] (h7);
\draw[thick,spinwave_total,cyan] (e7) to[out=45,in=135] (i7);

\draw[thick,spinwave_total,cyan] (c7) to[out=-45,in=-135] (g7);
\draw[thick,spinwave_total,cyan] (d7) to[out=-45,in=-135] (f7);

\foreach \n in {e7}
\node at (\n)[circle, fill, inner sep = 2pt, color = black]{};

\foreach \n in {c7,d7,f7,g7,h7,i7}
\node at (\n)[circle, fill, inner sep = 2pt, color = myred]{};

\coordinate[below = 4.25cm of b5, label = right:{$\!\!\!\!\!C_{35} = $}] (b8);
\coordinate[right = 1cm of b8, label = center:{$\dots$}] (c8);

\coordinate[below = 8cm of a2, label = {(b)}] (a9);

\coordinate[below = 0.3cm of a9, label = right:{$\!\!\!\!\!C_0$}] (b9);
\coordinate[right = 0.5cm of b9, label = center:{$=$}] (b9_2);
\coordinate[right = 0.4cm of b9_2] (c9);
\coordinate[right = 0.8cm of c9] (d9);
\coordinate[right = 0.8cm of d9] (e9);
\coordinate[right = 0.4cm of e9, label = center:{$+$}] (f9);

\draw[thick, double] (c9) -- (d9);
\draw[thick, double] (c9) -- (e9);

\draw[thick,spinwave_total,cyan] (c9) to[out=90,in=90] (d9);
\draw[thick,spinwave_total,cyan] (c9) to[out=90,in=90] (e9);

\foreach \n in {c9}
\node at (\n)[circle, fill, inner sep = 2pt, color = black]{};

\foreach \n in {d9,e9}
\node at (\n)[circle, fill, inner sep = 2pt, color = myred]{};

\coordinate[right = 0.4cm of f9] (a10);
\coordinate[right = 0.8cm of a10] (b10);
\coordinate[right = 0.8cm of b10] (c10);
\coordinate[right = 0.8cm of c10] (d10);
\coordinate[right = 0.8cm of d10] (e10);

\coordinate[below = 0.8cm of b10, label = {$F_C$}] (b10_low);

\draw[thick, double] (a10) -- (b10);
\draw[thick, double] (b10) -- (c10);
\draw[thick, double] (c10) -- (d10);
\draw[thick, double] (d10) -- (e10);

\draw[thick,spinwave_total,cyan] (b10) to[out=30,in=150] (d10);
\draw[thick,spinwave_total,cyan] (b10) to[out=45,in=135] (e10);

\draw[thick,spinwave_total,cyan,double] (a10) to[out=-30,in=-150] (c10);

\foreach \n in {a10,c10,d10,e10}
\node at (\n)[circle, fill, inner sep = 2pt, color = myred]{};

\foreach \n in {b10}
\node at (\n)[circle, fill, inner sep = 2pt, color = black]{};

\coordinate[below = 1.5cm of a9, label = {(c)}] (a14);

\coordinate[below = 0.3cm of a14, label = right:{$\!\!\!\!\!C\phantom{_0}$}] (b14);
\coordinate[right = 0.4cm of b14, label = center:{$=$}]    (c14);
\coordinate[right = 0.75cm of c14, label = center: {$C_0 \; + \; $}] (d14);

\coordinate[right = 0.5cm of d14]  (a15);
\coordinate[right = 0.7cm of a15]  (b15);
\coordinate[right = 0.35cm of b15, label=center:{$C_0$}] (B0_center_2);
\coordinate[right = 0.35cm of B0_center_2] (c15);
\coordinate[right = 0.7cm of c15] (d15);
\coordinate[right = 0.5cm of d15, label = center:{$\phantom{\dots}+\dots$}] (e15);

\draw[thick,double] (a15) -- (b15);
\draw (B0_center_2) circle (0.35cm);
\draw[thick,double] (c15) -- (d15);

\draw[spinwave_total,thick,cyan] (a15) to[out=-45,in=-135] (d15);

\foreach \n in {a15,d15}
\node at (\n)[circle, fill, inner sep = 2pt, color = myred]{};

\coordinate[right = 0.75cm of e15, label = center:{$=$}] (a16);
\coordinate[right = 0.75cm of a16, label = center: {$C_0 \; + \; $}] (b16);

\coordinate[right = 0.5cm of b16] (a17);
\coordinate[right = 0.7cm of a17] (b17);
\coordinate[right = 0.35cm of b17, label=center:{$C$}] (C_center);
\coordinate[right = 0.35cm of C_center] (c17);
\coordinate[right = 0.7cm of c17] (d17);

\draw[thick,double] (a17) -- (b17);
\draw (C_center) circle (0.35cm);
\draw[thick,double] (c17) -- (d17);

\draw[spinwave_total,thick,cyan] (a17) to[out=-45,in=-135] (d17);

\foreach \n in {a17,d17}
\node at (\n)[circle, fill, inner sep = 2pt, color = myred]{};

\end{tikzpicture}
\caption{Diagrammatic representation for the $C$ series. (a) The $C$ series comes from inserting the basic diagram, $C_{02}$, in the norm series, Fig. \ref{fig.B_series_terms}(a). The shown diagrams are the ones that give a nonzero contribution to the $C$ function. These come in an increasing order denoted $C_{n,n+2}$. At a given order $n$, there are $n$ interaction vertices to the left, $n+2$ to the right and a total of $n + 1$ nonzero diagrams. (b) Summation of all the last diagrams, $C_0 = \sum_{n} C_{n,n+2}^{(n)}$. The appearing $F_C$ function is shown in double spin wave lines. Diagrammatically, this looks identical to $F_B$. However, because the exterior is different, the structure of the self-consistent equation for $F_C$ [Eq. \eqref{eq.FC}] is different in this case. (c) The self-consistent equation for the full $C$ function is achieved by putting $0, 1, 2, \dots$ spin wave lines around $C_0$.}
\label{fig.C_series}
\end{figure}

The derivation of the $C$ function closely follows the derivation of the $B$ function in Sec. \ref{sec.derivation}. The basic diagram for the $C$ series is denoted $C_{02}$ and shown in Fig. \ref{fig.C_series}(a). Analogous to the $B$ series, the full $C$ series is then obtained by placing this basic diagram in the norm series, Fig. \ref{fig.B_series_terms}(a). This leads to the terms $C_{n,n+2}$, where $n$ is the number of spin waves from the adjoint state $\bra{\Psi_\bp}$ and $n+2$ is the number of spin waves from $\ket{\Psi_\bp}$. There are $n+1$ terms at order $n$: $C_{n, n+2} = \sum_{i = 0}^n C_{n, n + 2}^{(i)}$.
As for the $B$ series, we, then, first sum up only the last diagrams $C_{n,n+2}^{(n)}$ at each order, defining
\begin{equation}
C_0 = \sum_{n} C_{n,n+2}^{(n)}.
\label{eq.C_0_definition}
\end{equation}
This leads to the diagrammatic structure shown in Figs. \ref{fig.C_series}(b) and \ref{fig.C_series}(c). The result can be written in a similar form to $B_0$ and $F_B$, see Eqs. \eqref{eq.B0_and_FB} and \eqref{eq.FB}. Explicitly, 
\begin{align}
C_0(\bq_1, \bq_2; \bp, \omega) = N \big[ & g(\bp, \bq_1) g(\bp \!+\! \bq_1, \bq_2) G(\bp \!+\! \bq_1, \omega \!-\! \omega_{\bq_1}) \nn \\
 + & g(\bp, \bq_2) g(\bp \!+\! \bq_2, \bq_1) G(\bp \!+\! \bq_2, \omega \!-\! \omega_{\bq_2}) \big] \nn \\
& \cdot G(\bp \!+\! \bq_1 \!+\! \bq_2, \omega \!-\! \omega_{\bq_1} \!-\! \omega_{\bq_2}), \nn \\
& \cdot \left[1 + F_C(\bq_1, \bq_2; \bp, \omega)\right].
\label{eq.C0_and_FC_appendix}
\end{align}
Here, $F_C$ is given by Eq. \eqref{eq.FC}, while the above expression is identical to Eq. \eqref{eq.C0_and_FC}. By putting $0, 1, 2, \dots$ spin wave lines around $C_0$, as shown in Fig. \ref{fig.C_series}(c), the self-consistent equation for $C$ is achieved
\begin{align}
C(\bq_1, \bq_2; \bp, \omega) =&\; C_0(\bq_1, \bq_2; \bp, \omega) \nn \\
& + \sum_\bk g^2(\bp, \bk) G^2(\bp \!+\! \bk, \omega \!-\! \omega_\bk) \nn \\
& \cdot C(\bq_1, \bq_2; \bp \!+\! \bk, \omega \!-\! \omega_\bk),
\label{eq.self_consistency_equation_C}
\end{align}
which has the exact same structure as Eq. \eqref{eq.self_consistency_equation_B} for the $B$ function. 

\section{Vanishing diagrams} \label{app.vanishing_diagrams}
In this section, we show that all left-right asymmetric diagrams in the $B$ series vanish, and that the corresponding diagrams in the $C$ series vanish as well. \\

To understand how these diagrams vanish, we must first analyze a certain symmetry of the interaction vertex $g(\bp, \bk)$. Consider, therefore, a Bravais lattice in which the lattice points can be written
\begin{equation}
\br = n_1 \ba_1 + n_2 \ba_2 + n_3 \ba_3.
\label{eq.Bravais_lattice}
\end{equation}
Here, the $\ba_i$'s are the primitive vectors. In 2D, we simply set $n_3 = 0$. In 1D, $n_3 = n_2 = 0$. When the lattice is bi-partite, we can choose the primitive vectors to be nearest neighbors to a given site, such that $\bdelta \in \{\pm \ba_1, \pm \ba_2, \pm \ba_3\}$. Because of the periodic boundary conditions $\te^{i\bk \cdot \ba_j \cdot L_j} = 1$ for a $L_x\times L_y \times L_z$ lattice. In turn,
\begin{equation}
\bk \cdot \ba_j = n_j \cdot \frac{2\pi}{L_j}, 
\end{equation}
where $n_j = - L_j / 2 + 1, - L_j / 2 + 2, \dots, L_j / 2$ is an integer. Now, the wave vector of the antiferromagnetic spin-density wave $\bQ$ is defined by letting $n_j = L_j / 2$ for $j = x,y,z$. Then $\te^{i\bQ \cdot \ba_j} = \te^{i\pi} = -1$. In turn,
\begin{equation}
\gamma_{\bk + \bQ} = \frac{1}{z}\sum_{\bdelta} \te^{i \bQ\cdot \bdelta}\te^{i\bk \cdot \bdelta} = - \gamma_\bk,
\end{equation}
since $\te^{i\bQ\cdot\bdelta} = -1$, using that $\bdelta \in \{\pm \ba_1, \pm \ba_2, \pm \ba_3\}$. As in Appendix \ref{app.vanishing_diagrams}, the change in sign of $\gamma_\bk$ also means that the interaction changes sign
\begin{equation}
g(\bp, \bk + \bQ) = g(\bp + \bQ, \bk) = -g(\bp, \bk).
\label{eq.g_p_k_symmetry}
\end{equation}
On the other hand, the self-energy, and thereby the Green's function $G(\bp, \omega)$, is insensitive to this change in sign, because it scales with the square of the interaction. Therefore, 
\begin{equation}
G(\bp + \bQ, \omega) = G(\bp, \omega).
\label{eq.greens_function_symmetry}
\end{equation}
As we shall now show this leads to the vanishing of all asymmetric $B$ diagrams, as well as the corresponding diagrams in the $C$ series. 

\begin{figure}[t!]
\center
\begin{tikzpicture}[node distance=0.5cm and 1.5cm]
\coordinate (a1);
\coordinate[right = 0.75cm of a2, label = center:{$B_{2}^{\rm asym.} = $}] (b1);
\coordinate[right = 1cm of b1] (c1);
\coordinate[right = 1cm of c1] (d1);
\coordinate[right = 1cm of d1] (e1);
\coordinate[right = 1cm of e1] (f1);
\coordinate[right = 1cm of f1] (g1);

\draw[thick, double] (c1) -- (d1);
\draw[thick, double] (d1) -- (e1);
\draw[thick, double] (e1) -- (f1);
\draw[thick, double] (f1) -- (g1);

\draw[thick,spinwave_total,cyan] (c1) to[out=45,in=135] (e1);
\draw[thick,spinwave_total,cyan] (e1) to[out=90,in=90] (f1);

\draw[thick,spinwave_total,cyan] (d1) to[out=-30,in=-150] (g1);

\foreach \n in {e1}
\node at (\n)[circle, fill, inner sep = 2pt, color = black]{};

\foreach \n in {c1,d1,f1,g1}
\node at (\n)[circle, fill, inner sep = 2pt, color = myred]{};

\coordinate[below = 1.75cm of a1] (a2);
\coordinate[right = 0.75cm of a2, label = center:{$C_{13}^{\rm van.} = $}] (b2);
\coordinate[right = 1cm of b2] (c2);
\coordinate[right = 1cm of c2] (d2);
\coordinate[right = 1cm of d2] (e2);
\coordinate[right = 1cm of e2] (f2);
\coordinate[right = 1cm of f2] (g2);

\draw[thick, double] (c2) -- (d2);
\draw[thick, double] (d2) -- (e2);
\draw[thick, double] (e2) -- (f2);
\draw[thick, double] (f2) -- (g2);

\draw[thick,spinwave_total,cyan] (d2) to[out=60,in=120] (e2);
\draw[thick,spinwave_total,cyan] (d2) to[out=60,in=120] (g2);

\draw[thick,spinwave_total,cyan] (c2) to[out=-30,in=-150] (f2);

\foreach \n in {d2}
\node at (\n)[circle, fill, inner sep = 2pt, color = black]{};

\foreach \n in {c2,e2,f2,g2}
\node at (\n)[circle, fill, inner sep = 2pt, color = myred]{};

\end{tikzpicture}
\caption{Examples of vanishing $B$ and $C$ diagrams from orders $B_2$ and $C_{13}$ respectively.}
\label{fig.vanishing_diagrams}
\end{figure}
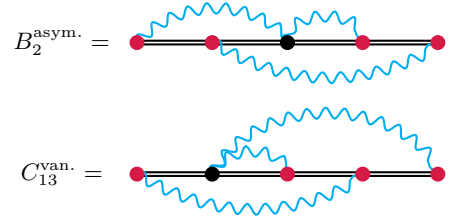

Specifically, the asymmetric diagrams are all of the form shown in the top of Fig. \ref{fig.vanishing_diagrams}. The example shown evaluates to
\begin{align}
B_2^{\rm asym.}(\bq_1, \bq_2; \bp, \omega) &= N \; g(\bp, \bq_2) G(\bp \!+\! \bq_2, \omega \!-\! \omega_{\bq_2}) \nn \\
& \cdot \sum_{\bk} g(\bp, \bk) G(\bp \!+\! \bk, \omega \!-\! \omega_\bk) \nn \\
& \cdot g(\bp \!+\! \bk,\bq_1) g(\bp \!+\! \bq_2, \bk) \nn \\
& \cdot G(\bp \!+\! \bk \!+\! \bq_1, \omega \!-\! \omega_\bk \!-\! \omega_{\bq_1}) \nn \\
& \cdot G(\bp \!+\! \bk \!+\! \bq_2, \omega\!-\! \omega_\bk \!-\! \omega_{\bq_2}). \!
\label{eq.B_2_asymmetric}
\end{align}
Here, the key point is that unlike the symmetric diagrams in Fig. \ref{fig.B_series_terms}, there is an \emph{odd} number of terms with interaction vertices $g(\cdot, \cdot)$ that depend on the summation index, $\bk$. It is, therefore, sensitive to changes in sign of $g$. As a result, the two terms $\bk$ and $\bk + \bQ$ in the sum in Eq. \eqref{eq.B_2_asymmetric} have the same magnitude, but opposite signs. Therefore, they cancel each other exactly. In this way, \emph{all} asymmetric diagrams in the $B$ series vanish identically. Another way to understand this vanishing is in terms of sublattice states. Every time the hole hops, it changes sublattice. Therefore, the asymmetric diagrams like the one shown in the top of Fig. \ref{fig.vanishing_diagrams} features overlaps of holes in opposite sublattices and thus vanish. Indeed, the symmetry in Eq. \eqref{eq.g_p_k_symmetry} of the interaction is due to the underlying sublattice symmetry of the system.

We finally comment on the vanishing of the remaining diagrams in the $C$ series. These diagrams have the structure shown in bottom part of Fig. \ref{fig.vanishing_diagrams}, where at least one spin wave line is between the external spin wave lines at momenta $-\bq_1$ and $-\bq_2$ joining the black dot. The reason that this vanishes is exactly the same as why the asymmetric diagrams in the $B$ series all vanish. Specifically, a sum $\sim \sum_\bk g(\bp, \bk)g(\bp + \bq_1, \bk)g(\bp + \bq_1 + \bk, \bq_2)$ appears, in which terms at $\bk$ and $\bk + \bQ$ cancel.

\section{Self-consistency equation for the scattering probability} \label{app.scattering_probability}
In this appendix, we derive a self-consistency equation for the scattering probability 
\begin{equation}
P^{\rm scat}_{\bp \to \bq} = \bra{\Psi_\bp}\hat{h}^\dagger_{\bq}\hat{h}_{\bq}\ket{\Psi_\bp} - Z_\bp \cdot \delta_{\bp, \bq}
\label{eq.scattering_probability_appendix}
\end{equation}
also defined in Eq. \eqref{eq.scattering_probability} of the main text. While this can be done diagrammatically as for the $B$ and $C$ functions, it is just as simple to write down the lowest order terms at a general energy $\omega$ and recognize the pattern. As for the $B$ and $C$ series, we omit the overall factor of the residue, $Z_\bp$, in the following. This must be included in the end. Since the final momentum must be $\bq$, the total change in crystal momentum is $\bq - \bp$. The lowest order term from the polaron wave function is thus $g^2(\bp, \bq - \bp) G^2(\bq, \omega - \omega_{\bq - \bp})$. As required, the scattering probability to a specific momentum state scales as $1 / N$, since $g^2(\cdot, \cdot) \propto 1 / N$. Inclusion of the next term yields
\begin{align}
P^{\rm scat}_{\bp \to \bq}(\omega) =\, & g^2(\bp, \bq - \bp) G^2(\bq, \omega - \omega_{\bq - \bp}) \nn \\
& + \sum_{\bk} g^2(\bp, \bk) G^2(\bp + \bk, \omega - \omega_\bk) \nn \\
\cdot & g^2(\bp + \bk, \bq - \bp - \bk) G^2(\bq, \omega - \omega_\bk - \omega_{\bq - \bp - \bk}) \nn \\
& + \dots \nn
\end{align}
In the second term, there is a single free momentum, $\bk$, but the final momentum must again be $\bq$. This gives the structure of the second term. This series continues indefinitely, but we notice that it can be rewritten as a self-consistency equation
\begin{align}
P^{\rm scat}_{\bp \to \bq}(\omega) =&\, g^2(\bp, \bq - \bp) G^2(\bq, \omega - \omega_{\bq - \bp}) \nn \\
&\, + \sum_{\bk} g^2(\bp, \bk) G^2(\bp + \bk, \omega - \omega_\bk) \nn \\
&\, \cdot P^{\rm scat}_{\bp + \bk \to \bq} (\omega - \omega_\bk).
\label{eq.scattering_probability_self_consistent}
\end{align}
Like the equations for the $B$- and $C$-functions, this has the exact same structure as the equation for the derivative of the self-energy [Eq. \eqref{eq.derivative_self_energy}]. By solving this iteratively, evaluating at the quasiparticle peak, $\varepsilon_\bp$, and multiplying the result by the residue $Z_\bp$, we obtain the scattering probabilities plotted in Fig. \ref{fig.magnetization_vs_scattering_probability}. If we sum up all contributions in Eq. \eqref{eq.scattering_probability_appendix}, we obtain $\sum_{\bq} P^{\rm scat}_{\bp \to \bq} = 1 - Z_\bp$, as one might expect. This summation rule gives a good consistency check for the numerical calculations. 

\begin{figure}[tb]
	\begin{center}
		\includegraphics[width=\columnwidth]{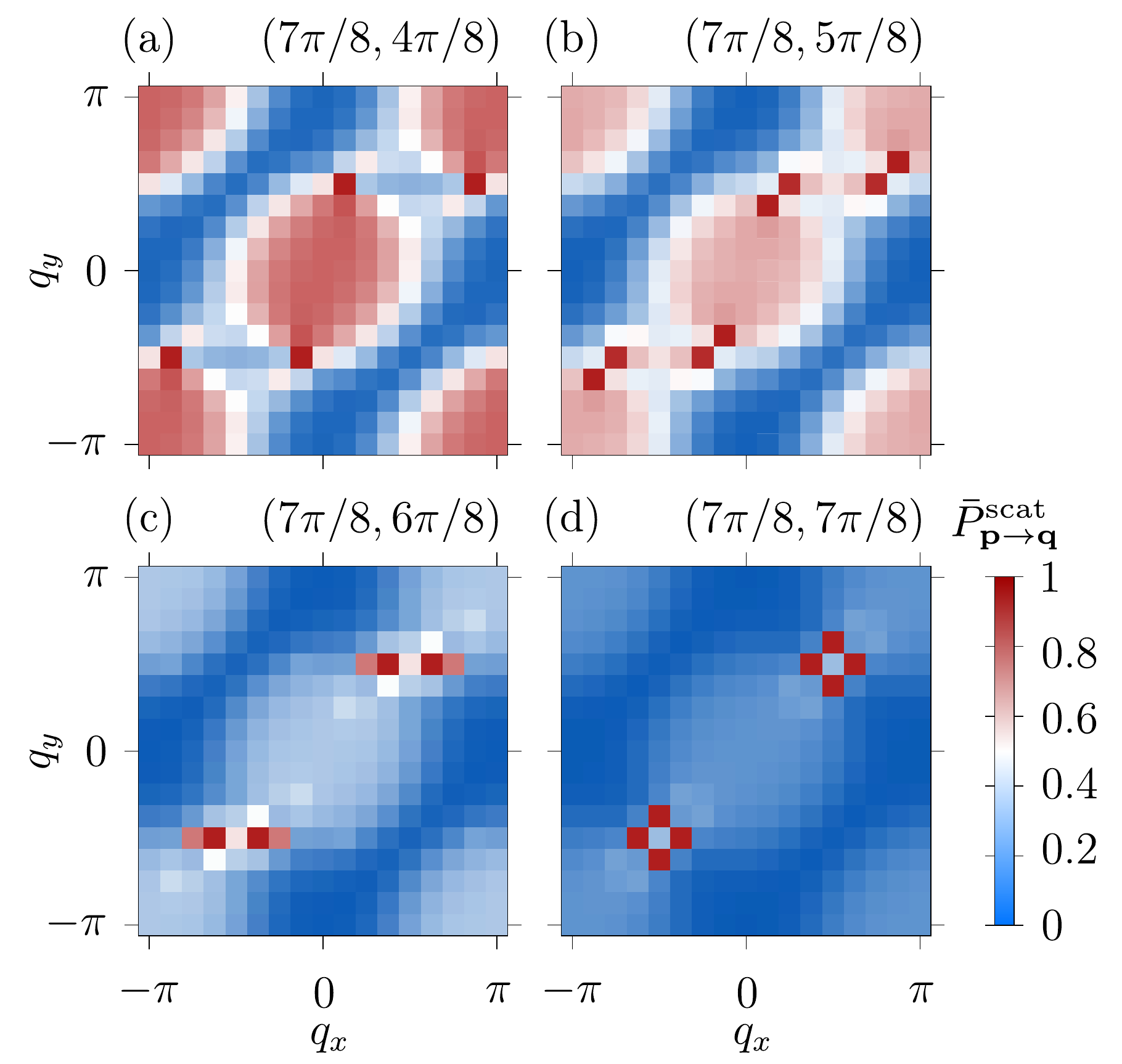}
	\end{center}
	\caption{$C_2$ symmetrized scattering probability $\bar{P}^{\rm scat}_{\bp \to \bq} = P^{\rm scat}_{\bp \to +\bq} + P^{\rm scat}_{\bp \to -\bq}$ normalized to their maximal values for various indicated crystal momenta and an interaction strength $J / t = 0.3$ in a 16 by 16 square lattice. Moving through panels (a) to (d), we increase $p_y$ from $4\pi / 8$ to $7\pi / 8$, keeping $p_x = 7\pi / 8$ constant. Panel (a) corresponds to Fig. \ref{fig.magnetization_vs_scattering_probability}(e). }
	\label{fig.scattering_profiles_examples} 
\end{figure} 

In Figs. \ref{fig.magnetization_vs_scattering_probability}(d)--\ref{fig.magnetization_vs_scattering_probability}(e), we noticed that the scattering profiles are remarkably sharp. We look into this in Fig. \ref{fig.scattering_profiles_examples}, where we vary $p_y$ from $4\pi / 8$ to $7\pi / 8$, keeping a constant $p_x = 7\pi / 8$ as in Fig. \ref{fig.magnetization_vs_scattering_probability}(e). This shows that the sharp scattering profiles in Figs. \ref{fig.magnetization_vs_scattering_probability}(d)--\ref{fig.magnetization_vs_scattering_probability}(e) are by no means exceptional. In fact, at $\bp = (7\pi / 8, 7\pi / 8)$, the scattering probability is sharpened further, and the hole dominantly scatters to only eight distinct momentum states. 

\section{Analytical result in the Ising limit} \label{app.ising_limit}
In this appendix, we derive an analytical formula for the magnetization in the Ising limit, $\alpha = 0$. This is possible due to a huge simplification in the interaction, $g(\bq, \bk) \to zt \cdot \gamma_{\bq + \bk} / \sqrt{N}$. As a result of this simplification, the Green's function within SCBA is independent of crystal momentum, and fulfills the equation $G^{-1}(\omega) = \omega - zJ / 2 - zt^2 \cdot G(\omega \!-\! zJ/2)$. This allows for an analytical solution in terms of a continued fraction \cite{Starykh1996}
\begin{equation}
G(\omega) = \frac{1}{\omega \!-\! zJ / 2 \!-\! zt^2 \frac{1}{\omega - 2 \cdot zJ / 2 - zt^2 \frac{1}{\omega - 3 \cdot zJ / 2 \dots}} },
\label{eq.hole_propagator_ising_limit}
\end{equation}
facilitating a numerically simple implementation. We now show that a similar description applies to the magnetization. 

Since the antiferromagnetic coherence factors in the Ising limit are $u_\bk \to 1$ and $v_\bk \to 0$, the $C$ series does not contribute to the magnetization [Eq. \eqref{eq.C_position_space}]. Focusing then on the $B$ series, we first calculate $F_B$ from Eq. \eqref{eq.FB}
\begin{align}
F_B(\bq_1 \!-\! \bq_2; \omega) =& \, zt^2 \gamma_{\bq_1 - \bq_2} G^2(\omega \!-\! zJ) \nn \\
& \cdot \left[1 + F_B(\bq_1 \!-\! \bq_2, \omega \!-\! zJ / 2)\right] \nn \\
=& \sum_{n = 1}^\infty (zt^2 \gamma_{\bq_1 - \bq_2})^n \prod_{k = 1}^n G^2\!\left(\omega \!-\! (k + 1)\frac{zJ}{2}\right).
\label{eq.F_B_analytical_result_ising_limit}
\end{align}
To obtain the upper line, we assume that $F_B$ is independent of the polaron crystal momentum $\bp$, and use that $\sum_{\bk} \gamma_{\bk + \bq_1} \gamma_{\bk + \bq_2} / N = \gamma_{\bq_1 - \bq_2} / z$. Note that $F_B$ only depends on relative momentum, $\bq_1 - \bq_2$. Then, by repeatedly reinserting $F_B$ as described by the upper line, we obtain the infinite series in the lower line. From Eq. \eqref{eq.B0_and_FB}, it then follows that 
\begin{align}
B_0(\bq_1, \bq_2; \bp, \omega) =& \; (zt)^2 \gamma_{\bp + \bq_1} \gamma_{\bp + \bq_2} G^2(\omega\! -\! zJ / 2)\nn \\
& \cdot \left[1 + F_B(\bq_1 \!-\! \bq_2; \omega)\right].
\label{eq.B_0_ising_limit}
\end{align}
This, therefore, still depends on the momentum of the polaron, $\bp$. To get rid of this momentum dependency in the self-consistency equations, we define
\begin{align}
\!\!\!\!\!\Delta B(\bq_1, \bq_2; \omega) = B(\bq_1, \bq_2; \bp, \omega) - B_0(\bq_1, \bq_2; \bp, \omega), \!\!
\label{eq.B_tilde_ising_limit}
\end{align}
and make the ansatz that this is independent of $\bp$. Using Eq. \eqref{eq.self_consistency_equation_B}, we, in fact, get
\begin{align}
\Delta B(\bq_1, \bq_2; \omega) =& \; \Delta B_0(\bq_1, \bq_2; \omega) \!+\! zt^2 \cdot G^2(\omega \!-\!zJ / 2) \nn \\
& \cdot \Delta B(\bq_1, \bq_2; \omega \!-\! zJ / 2), 
\label{eq.self_consistency_equation_Delta_B_ising} 
\end{align}
where 
\begin{align}
\Delta B_0(\bq_1, \bq_2; \omega) =& (zt^2)^2 \Big[2\gamma_{\bq_1}\gamma_{\bq_2} + \left(1 - \frac{2}{z}\right) \gamma_{\bq_1 - \bq_2} \nn \\
& - \frac{\gamma_{\bq_1 + \bq_2}}{z} \Big] G^2(\omega\! -\! zJ / 2) G^2(\omega \!-\! zJ) \nn \\
&\cdot \left[1 + F_B(\bq_1 \!-\! \bq_2; \omega \!-\! zJ / 2)\right]
\label{eq.Delta_B_0_ising} 
\end{align}
Here, we use that $\sum_\bk \gamma_{\bk}^2 \gamma_{\bk + \bq_1} \gamma_{\bk + \bq_2} / N = [2\gamma_{\bq_1}\gamma_{\bq_2} + (1 - 2 / z) \gamma_{\bq_1 - \bq_2} - \gamma_{\bq_1 + \bq_2} / z] / z^2$. In the final step, we repeatedly insert $\Delta B$ on the right-hand side, starting from the initial value of $\Delta B = 0$. Thus,
\begin{align}
&\Delta B(\bq_1, \bq_2; \omega) = \Delta B_0(\bq_1, \bq, \omega) \!+\! \sum_{n = 1}^{\infty} (zt^2)^n \nn \\
& \cdot \prod_{k=1}^n \!G^2\!\left(\omega - k \frac{zJ}{2}\right) \Delta B_0\!\left(\bq_1, \bq_2; \omega \!-\! n \frac{z J}{2}\right).
\label{eq.analytical_result_B_tilde_ising_limit}
\end{align}
To get the sought analytical result for the $B$ series, we then combine Eqs. \eqref{eq.analytical_result_B_tilde_ising_limit} for $\Delta B$ and \eqref{eq.B_tilde_ising_limit} for $B$ with Eq. \eqref{eq.Delta_B_0_ising} for $\Delta B_0$, Eq. \eqref{eq.B_0_ising_limit} for $B_0$, and Eq. \eqref{eq.F_B_analytical_result_ising_limit} for $F_B$. In all of these expressions, we use the continued fraction form for the Green's function in Eq. \eqref{eq.hole_propagator_ising_limit}. Finally, the magnetization in the neigborhood of the hole is achieved by multiplying $B$ with the (momentum independent) residue $Z$ and then transforming to position space using Eqs. \eqref{eq.spin_magnetization_2} and \eqref{eq.B_position_space}.

Since $B_0$ depends explicitly on the polaron crystal momentum, $\bp$, we might expect that the magnetization in position space will as well. We show now, however, that the magnetization becomes independent of the polaron momentum just as the Green's function [Eq. \eqref{eq.hole_propagator_ising_limit}]. The magnetization, $M(\bd) = \Delta M(\bd) + M_0(\bp, \bd)$, can be separated into two terms using Eq. \eqref{eq.B_tilde_ising_limit}
\begin{align}
\Delta M(\bd) &= \frac{1}{N^2} \sum_{\bq_1, \bq_2} \te^{i(\bq_1 - \bq_2)\cdot\bd} \Delta B(\bq_1, \bq_2; \varepsilon_0)\nn \\
M_0(\bp, \bd) &= \frac{1}{N^2} \sum_{\bq_1, \bq_2} \te^{i(\bq_1 - \bq_2)\cdot\bd} B_0(\bq_1, \bq_2; \bp, \varepsilon_0). 
\label{eq.tilde_M_and_M_0_ising}
\end{align}
Here, we explicitly evaluate the functions at the quasiparticle ground state energy, $\varepsilon_0 = \Sigma(\varepsilon_0)$. While the first term, $\Delta M(\bd)$, is explicitly independent of $\bp$, the second term, $M_0(\bp, \bd)$, might still depend on $\bp$. We now insert Eq. \eqref{eq.B_0_ising_limit}, writing $\Delta \bq = \bq_1 - \bq_2$ and using that $B_0(\bq_1, \bq_2; \bp, \varepsilon_0) = Z \gamma_{\bp + \bq_1} \gamma_{\bp + \bq_2} b_0(\Delta\bq; \varepsilon_0)$. Here, $b_0(\Delta\bq; \varepsilon_0) = (zt)^2 G^2(\omega\! -\! zJ / 2) \left[1 + F_B(\Delta\bq; \omega\! -\! zJ / 2)\right]$. Consequently, we get
\begin{align}
& M_0(\bp, \bd) = \frac{Z}{N^2}\!\! \sum_{\Delta\bq, \bq_1} \te^{i\Delta \bq \cdot \bd}b_0(\Delta\bq; \varepsilon_0) \gamma_{\bp + \bq_1}\gamma_{\bp + \bq_1 - \Delta\bq} \nn \\ 
&= \frac{Z}{(zN)^2}\!\! \sum_{\Delta\bq, \bq_1}\!\!\! \te^{i\Delta \bq \cdot \bd} b_0(\Delta\bq; \varepsilon_0)\! \sum_{\bdelta_1, \bdelta_2} \!\! \te^{i(\bp + \bq_1)\cdot(\bdelta_1 + \bdelta_2) -i\Delta\bq \cdot \bdelta_2}. \nn 
\end{align}
Note that $Z$ is the quasiparticle residue, while $z$ is the coordination number. The sum over $\bq_1$ now enforces $\bdelta_2 = - \bdelta_1$, 
\begin{align}
M_0(\bd) =& \, \frac{Z}{z^2 N} \sum_{\Delta\bq} \te^{i\Delta \bq \cdot \bd}b_0(\Delta\bq; \varepsilon_0) \sum_{\bdelta_1} \te^{-i\Delta\bq \cdot \bdelta_1} \nn \\
=& \, \frac{Z}{zN} \sum_{\Delta\bq} \te^{i\Delta \bq \cdot \bd}b_0(\Delta\bq; \varepsilon_0) \gamma_{\Delta \bq}. 
\label{eq.M_0_final_ising}
\end{align}
This shows explicitly that the magnetization is independent of the polaron momentum, $\bp$, in the Ising case.

\section{Symmetries of the $B$ and $C$ functions} \label{app.symmetries_of_B_and_C}
From the equations for the $B$ and $C$ function [Eqs. \eqref{eq.B_position_space} and \eqref{eq.C_position_space}]
\begin{align}
\!\!\!\! B(\bq_1, \bq_2; \bp, \omega) &= N \sum_\bk \braket{\hat{h}^\dagger_{\bk + \bq_1} \hat{h}_{\bk + \bq_2} \hat{b}^\dagger_{-\bq_1} \hat{b}_{-\bq_2}}_{\bp, \omega}, \nn \\
\!\!\!\! C(\bq_1, \bq_2; \bp, \omega) &= N \sum_{\bk} \braket{\hat{h}^\dagger_{\bk - \bq_1 - \bq_2}\hat{h}_\bk \hat{b}_{-\bq_1} \hat{b}_{-\bq_2}}_{\bp, \omega}. \!
\label{eq.B_and_C_position_space_appendix}
\end{align}
and the spatial symmetries of the interactions, it follows that the $B$ and $C$ functions have 3 essential symmetry properties ($X = B, C$)
\begin{enumerate}
	\item Total exchange symmetry: \\ $X(\bq_2, \bq_1; \bp, \omega) = X(\bq_1, \bq_2; \bp, \omega)$
	\item Any exchange of $(x, y, z)$ coordinates: \\ $X(\bq_1^*, \bq_2^*; \bp^*, \omega) = X(\bq_1, \bq_2; \bp, \omega)$. \\ Example: $\bq^* = (q_x, q_y, q_z)^* = (q_y, q_x, q_z)$. 
	\item Sign flip of individual coordinates: \\ $X(\bar{\bq}_1, \bar{\bq}_2; \bar{\bp}, \omega) = X(\bq_1, \bq_2; \bp, \omega)$. \\ Example $\bar{\bq} = (-q_x, q_y, q_z)$. 
\end{enumerate}
For the $C$ function, the total exchange symmetry (1.) follows directly from the symmetric form of $C$ in Eq. \eqref{eq.B_and_C_position_space_appendix}. For the $B$ function, we additionally use that it is real so that $B(\bq_1, \bq_2; \bp, \omega) = B^*(\bq_1, \bq_2; \bp, \omega) = N \sum_\bk \braket{(\hat{h}^\dagger_{\bk + \bq_1} \hat{h}_{\bk + \bq_2} \hat{b}^\dagger_{-\bq_1} \hat{b}_{-\bq_2})^\dagger}_{\bp, \omega} = B(\bq_2, \bq_1; \bp, \omega)$. The reality of $B(\bq_1, \bq_2; \bp, \omega)$ is a result of the fact that all coefficients in the polaron wave function expansion in Eq. \eqref{eq.Reiters_wave_function} are real. This, in turn, is a consequence of the fact that all appearing Green's functions are always evaluated \emph{below} the quasiparticle peak. The second and third symmetries reflect that there is no preferred direction of the system. Therefore, we can swap the coordinates as we wish (2.) and reverse 1, 2 or all 3 spatial directions (3.).

\bibliographystyle{apsrev4-1}
\bibliography{ref_magnetic_polaron}

\end{document}